\documentclass[lettersize,journal]{IEEEtran}
\usepackage{amsmath,amsfonts}
\usepackage{algorithmic}
\usepackage{algorithm}
\usepackage{array}
\usepackage[caption=false,font=normalsize,labelfont=sf,textfont=sf]{subfig}
\usepackage{textcomp}
\usepackage{stfloats}
\usepackage{url}
\usepackage{verbatim}
\usepackage{graphicx}
\usepackage{cite}
\usepackage[colorlinks=true, citecolor=blue, urlcolor=blue, linkcolor=blue]{hyperref} 
\usepackage{multirow} 
\usepackage{threeparttable} 
\usepackage{setspace} 
\usepackage{booktabs} 
\newcommand{\tabincell}[2]{\begin{tabular}{@{}#1@{}}#2\end{tabular}} 
\usepackage{pmboxdraw} 
\usepackage{amsthm} 
\theoremstyle{definition}

\usepackage{makecell} 

\usepackage{orcidlink} 

\usepackage{xcolor, soul} 
\usepackage{CJKutf8} 

\sethlcolor{yellow} 

\begin{document}
\begin{CJK}{UTF8}{gbsn} 
	
\title{
	{\fontsize{24}{28}\selectfont\spaceskip=0.35em{A High Performance and Efficient Post-Quantum Crypto-Processor for FrodoKEM}}
}
	
\author{
	Kai Li~\orcidlink{0009-0008-3824-4672}, Jiahao Lu, Fu Yao, Guang Zeng, Dongsheng Liu, Shengfei Gu, Zhengpeng Zhao, Jiachen Wang
	
	\thanks{
		This work has been submitted to the IEEE for possible publication. Copyright may be transferred without notice, after which this version may no longer be accessible.
	}
}

\markboth{IEEE XXX,~VOL.~XX, No.~XX, January~2026}
{Shell \MakeLowercase{\textit{et al.}}: A Sample Article Using IEEEtran.cls for IEEE Journals}

\maketitle

\begin{abstract}

FrodoKEM is a lattice-based post-quantum key encapsulation mechanism (KEM). It has been considered for standardization by the International Organization for Standardization (ISO) due to its robust security profile.
However, its hardware implementation exhibits a weakness of high latency and heavy resource burden, hindering its practical application. Moreover, diverse usage scenarios call for comprehensive functionality.
To address these challenges, this paper presents a high-performance and efficient crypto-processor for FrodoKEM.
A multiple-instruction overlapped execution scheme is introduced to enable efficient multi-module scheduling and minimize operational latency.
Furthermore, a high-speed, reconfigurable parallel multiplier array is integrated to handle intensive matrix computations under diverse computation patterns, significantly enhancing hardware efficiency. 
In addition, a compact memory scheduling strategy shortens the lifespan of intermediate matrices, thereby reducing overall storage requirements.
The proposed design provides full support for all FrodoKEM security levels and protocol phases. 
It consumes 13467 LUTs, 6042 FFs, and 14 BRAMs on an Artix-7 FPGA and achieves the fastest reported execution time.
Compared with state-of-the-art hardware implementations, our design improves the area-time product (ATP) by 1.75$\times$--2.00$\times$.

\end{abstract}

\begin{IEEEkeywords}
	Post-quantum cryptography(PQC), lattice-based cryptography, FrodoKEM, compact memory scheduling, reconfigurable multiplier array
\end{IEEEkeywords}

\section{Introduction}

\IEEEPARstart{T}{he} rapid advancement of quantum computing equipped with Shor's algorithm~\cite{Shor_1994_Algorithms,Shor_1999_Polynomial} poses a fundamental threat to traditional public key infrastructure.
Post-quantum cryptography (PQC) addresses this challenge by developing cryptographic primitives that remain secure against both classical and quantum adversaries.
Lattice-based cryptography~\cite{Ajtai_1996_Generating,Ajtai_1997_Publickey} is a leading approach in PQC.
Among its hardness assumptions, the Learning With Errors (LWE) problem~\cite{Regev_2004_Quantum} underlies a wide range of cryptographic schemes.
The International Organization for Standardization (ISO) has taken concrete steps toward post-quantum cryptography standardization.
In April 2023, ISO/IEC agreed to include FrodoKEM in the revision of ISO/IEC~18033-2.

FrodoKEM~\cite{Bos_2016_Frodo} is an LWE-based key encapsulation mechanism (KEM) defined over algebraically unstructured lattices.
Unlike widely deployed post-quantum schemes such as Kyber, which are based on structured lattices, it avoids ring and module constructions.
Its security relies directly on the standard LWE problem and follows a conservative approach.
This results in a complete and well-established security proof and provides strong resistance against both classical and quantum attacks.
Consequently, FrodoKEM has been endorsed for post-quantum cryptographic migration by national security agencies, including the German Federal Office for Information Security (BSI) and the French National Agency for the Security of Information Systems (ANSSI)~\cite{Urbano_2025_FrodoKEM}.

Existing software implementations focus on accelerating matrix operations and reducing memory usage.
Assembly-level multiplication routine has been developed to improve performance and reduce peak stack usage in~\cite{Howe_2018_Standard}.
Bos \emph{et al.}~\cite{Bos_2018_Fly} leverage the single-instruction multiple-data (SIMD) paradigm to accelerate matrix operations.
On ARMv8 processors, vector registers and vector instructions are used to implement parallel matrix multiplication~\cite{Kwon_2021_ARMed}.
Despite these optimizations, software implementations still exhibit high execution latency.
GPUs have also been used to parallelize matrix operations~\cite{An_2020_Efficient,Gupta_2021_PQC,Lee_2022_DPCrypto,Lee_2022_Efficient}, which achieve high throughput but require powerful computing platforms.

Software/hardware (SW/HW) co-design improves performance while retaining software flexibility by offloading compute-intensive tasks to hardware.
The Sapphire crypto-processor~\cite{Banerjee_2019_Sapphire_ext} achieves high efficiency and low power, but adopts a tiling strategy that differs from the official standard.
A RISC-V-based co-design~\cite{Karl_2022_Hardware} integrates tightly coupled hardware accelerators and supports all parameter sets. Nevertheless, it still requires tens of millions of cycles to complete execution.
Dang \emph{et al.}~\cite{Dang_2019_Implementing} implement matrix generation and multiplication entirely in hardware. They reduce peak execution time to only a few milliseconds. However, each security level is implemented separately.

Hardware implementations target high performance under constrained hardware resources. 
Early FPGA designs~\cite{Howe_2018_Standard} were efficiently implemented on \mbox{Artix-7} devices with low logic and memory overhead.
Their subsequent works~\cite{Howe_2019_Optimised} further improved throughput by parallelizing matrix multiplication and replacing SHAKE with a lightweight pseudo-random number generator (PRNG), namely Trivium. 
The use of Trivium reduces hardware cost but deviates from the standard specification and may raise security concerns. 
Moreover, the above implementations target early NIST Round~1 submissions and differ from the latest standard specification~\cite{Urbano_2025_FrodoKEM}.
More recently, Duzyol \emph{et al.}~\cite{Duzyol_2025_Can} proposed a fully standard-compliant architecture.
Their design scales throughput by increasing the number of DSPs and BRAMs.
It achieves 976--1077 operations per second for FrodoKEM-640.
Nevertheless, only the lowest security level is supported. The high resource consumption further limits overall efficiency, with a single protocol phase requiring 18--20 BRAM blocks.
High-Level Synthesis (HLS) has also been explored. However, current HLS designs still suffer from limited performance~\cite{Basu_2019_NIST,Urbano_2025_FrodoKEM}. 

It can be observed that high-performance implementations of FrodoKEM remain rare. 
Existing pure hardware designs suffer from long execution latency and low efficiency. 
Most of them target only specific security levels or protocol phases, which limits their versatility across scenarios. 
HW/SW co-design approaches provide flexibility at the cost of performance. 
Thus, there is a pressing need for a hardware solution that delivers high performance and efficiency while maintaining configurability.

This work proposes a high-speed, efficient, and reconfigurable post-quantum cryptographic processor for FrodoKEM. The contributions are summarized as follows:

\begin{itemize}
	\item 
	A multiple-instruction overlapped execution scheme is developed to enable overlapped execution of independent operations.
	This approach achieves at least a $1.65\times$ speedup.	
	
	\item 
	To handle intensive matrix computation demands across versatile operation patterns, a high-speed and reconfigurable parallel multiplier array is devised.
	By exploiting sign extraction, the design eliminates the need for DSP blocks, significantly enhancing hardware efficiency.
	
	\item 
	A compact memory scheduling strategy is employed to mitigate storage overhead.
	By shortening the lifespan of intermediate variables, BRAM usage is reduced to 14, which is 30\% less than prior designs~\cite{Duzyol_2025_Can}.
	Interleaved and ping-pong access techniques are adopted to sustain high throughput.
	
	\item 
	The proposed processor provides comprehensive support for all FrodoKEM security levels and protocol phases.
	Implemented on an Artix-7 FPGA, the design consumes 6609 equivalent slices.
	Compared with state-of-the-art implementations, it achieves a $1.75\times$--$2.00\times$ improvement in area-time product (ATP) and a 6\%-16\% execution time reduction.
\end{itemize}

The remainder of this paper is organized as follows. \mbox{Section}~\ref{sec:Preliminary} illustrates FrodoKEM algorithm. \mbox{Section}~\ref{sec:Hardware} describes the detailed hardware architecture of the designed crypto-processor. \mbox{Section}~\ref{sec:Results} presents the implementation results, followed by a discussion of the performance compared to related works. Finally, this paper is concluded in \mbox{Section}~\ref{sec:Conclusion}.

\section{Preliminary} \label{sec:Preliminary}

In this section, we present the necessary background knowledge on the FrodoKEM algorithm to lay the foundation for the subsequent hardware implementation chapters.

\subsection{Notations}

The notations used in this paper are based on the standard specification~\cite{frodokem_2025}. Let $q = 2^{D}$ be a power-of-two modulus, where $D \in \{15,16\}$ depending on the security level. The ring of integers modulo $q$ is denoted by $\mathbb{Z}_q$. Matrices over $\mathbb{Z}_q$ are represented by bold uppercase letters (e.g., $\mathbf{A}$). For a matrix $\mathbf{A} \in \mathbb{Z}_q^{n \times m}$,  $a_{i,j}$ denote the element in the $i$-th row and $j$-th column of a matrix, where $i,j$ start from $0$. The transpose of $\mathbf{A}$ be denoted by $\mathbf{A}^{T}$. When a matrix $\mathbf{A}$ is partitioned into blocks, $\mathbf{A}^{(b_r \times b_c)}$ denotes a block of $\mathbf{A}$ with size $b_r \times b_c$. Accordingly, the block located at the $u$-th block row and $v$-th block column is denoted by $\mathbf{A}^{(b_r \times b_c)}_{u,v} \in \mathbb{Z}_q^{b_r \times b_c}$. In the special case where the matrix is partitioned only along the row dimension, i.e., $b_c = n$, the $u$-th row block is denoted by $\mathbf{A}^{(b_r \times \cdot)}_{u}$. Similarly, when the matrix is partitioned only along the column dimension, the $v$-th column block is denoted by $\mathbf{A}^{(\cdot \times b_c)}_{v}$.

The error distribution $\chi$ is a symmetric discrete distribution over $\mathbb{Z}$, parameterized by a cumulative distribution table $T_{\chi}$ as defined in the specification~\cite{frodokem_2025}.

Bit strings are concatenated using the operator $\|$. For structured objects such as matrices, concatenation implies that the elements are first serialized into bit strings and then concatenated.

\subsection{LWE-based Matrix Operations}

Unlike structured lattice-based schemes (e.g., Kyber) that utilize the Number Theoretic Transform (NTT) for polynomial multiplication, FrodoKEM is built upon the standard Learning with Errors (LWE) problem. As a result, its dominant computational primitive is large-scale dense matrix--matrix multiplication. As shown in~(\ref{eq:lwe_core}), the core operation involves
\begin{equation}
	\label{eq:lwe_core}
	\mathbf{B} = \mathbf{A}\mathbf{S} + \mathbf{E} \pmod{q},
\end{equation}
where $\mathbf{A} \in \mathbb{Z}_q^{n \times n}$, 
$\mathbf{S} \in \mathbb{Z}_q^{n \times \bar{n}}$, and 
$\mathbf{E} \in \mathbb{Z}_q^{n \times \bar{n}}$ denote the public, secret, and error matrices, respectively. 
The parameter $n$ represents the dimension of the square public matrix, while $\bar{n}$ denotes an auxiliary dimension associated with the secret and error matrices.

From a hardware implementation perspective, a notable advantage is offered: since $q$ is a power of two, modular reduction can be implemented by retaining the lower $D$ bits, thereby eliminating the need for costly Montgomery or Barrett reduction units. However, the matrix dimensions are extremely large (up to $n=1344$), leading in a computational complexity of $\mathcal{O}(n^2\bar{n})$. Moreover, the public matrix $\mathbf{A}$ requires several megabytes of storage, which makes full storage impractical on FPGA platforms. Consequently, $\mathbf{A}$ is generated row-by-row and buffered during matrix multiplication, rather than being fully stored.

\subsection{Matrix Generation and Error Sampling}

FrodoKEM relies on cryptographic hash functions and discrete Gaussian sampling to generate the public matrix and error matrices.
These operations are performed by the functions \textsf{Gen} and \textsf{Sample}. Both \textsf{Gen} and \textsf{Sample} employ extendable-output functions from the SHA-3 family~\cite{fips_202}, namely \mbox{SHAKE128} and \mbox{SHAKE256}.
\mbox{SHAKE} denotes either \mbox{SHAKE128} or \mbox{SHAKE256}, depending on the security level: \mbox{SHAKE128} is used for Level~1 (FrodoKEM-640), while \mbox{SHAKE256} is used for Level~3 and Level~5 (FrodoKEM-976 and FrodoKEM-1344).
\mbox{SHAKE} is built on the KECCAK permutation, which operates on a 1600-bit internal state with 24-round iterations.
After absorbing specific seeds, it produces pseudo-random bit streams of arbitrary length and serves as the primary source of randomness.

According to the standard specification~\cite{frodokem_2025}, the function \textsf{Gen} deterministically generates the public matrix $\mathbf{A}$ from the seed $seed_{\mathbf{A}}$ using either AES128 or \mbox{SHAKE128}. In this work, \mbox{SHAKE128} is adopted to provide stronger security and to enable a unified hash-based implementation. The expanded output stream is parsed to construct the entries of $\mathbf{A}$ over $\mathbb{Z}_q$. The function \textsf{Sample} generates the secret matrix $\mathbf{S}$ and the other error matrices (e.g., $\mathbf{E}$, $\mathbf{S}' \in \mathbb{Z}_q^{\bar{n} \times n}$, $\mathbf{E}' \in \mathbb{Z}_q^{\bar{n} \times n}$ and
$\mathbf{E}'' \in \mathbb{Z}_q^{\bar{n} \times \bar{n}}$). It produces samples from the discrete error distribution $\chi$ by mapping uniformly distributed values derived from \mbox{SHAKE} to integer error samples using a cumulative distribution function (CDF) based sampling method. The maximum absolute value of the samples is bounded by the parameter $d$. In addition, \mbox{SHAKE} is also used for seed derivation and hash digest generation.

\subsection{Encoding and Decoding Functions}

The functions \textsf{Encode} and \textsf{Decode} implement the conversion between a binary message and a $\bar{n} \times \bar{n}$ matrix with elements over $\mathbb{Z}_q$. In this mapping, each group of $B$ message bits corresponds to a single matrix entry, where the encoding bit width $B \in \{2,3,4\}$ depends on the selected security level. The binary message is typically denoted by $u$ and has a length of $\bar{n}^2 \cdot B$ bits. In all parameter sets, $\bar{n}$ is fixed to $8$, resulting in a message length of $64B$ bits. The function \textsf{Encode} maps each $B$-bit block of the message to an element in $\mathbb{Z}_q$ by scaling it by a factor of $q/2^B$, thereby forming an encoded matrix representation. Conversely, \textsf{Decode} performs the inverse operation by rounding the matrix elements and recovering the corresponding $B$-bit message blocks. Since the decoding procedure involves a rounding operation, \textsf{Decode} is not a strict inverse of \textsf{Encode}.

\subsection{Algorithm Description}

\begin{figure}[t]
	\centering
	\includegraphics[width=0.95\linewidth]{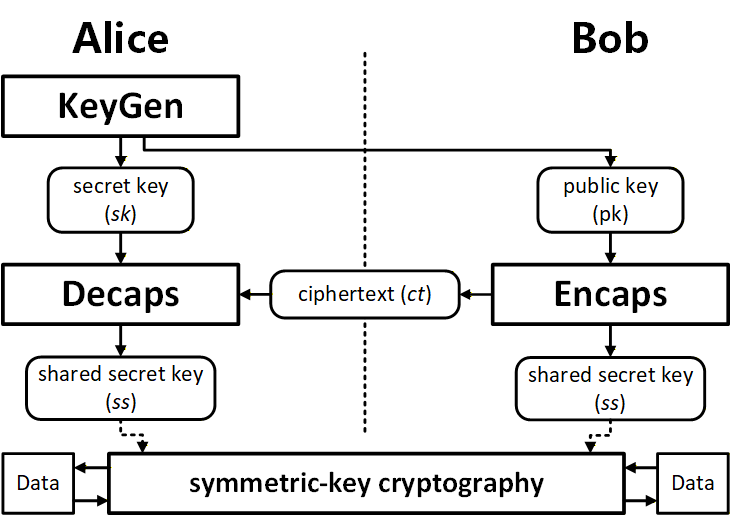}
	\caption{Block diagram of key encapsulation mechanism (KEM).}
	\label{fig:KEM}
\end{figure}

A key encapsulation mechanism (KEM) is formally defined by a tuple of three phases, namely \textit{KeyGen}, \textit{Encaps}, and \textit{Decaps}, as illustrated in Fig.~\ref{fig:KEM}:

\begin{itemize}
	\item \textit{KeyGen}()$\overset{\$}{\rightarrow}$($pk$, $sk$): A key generation algorithm that generates a public key $pk$ and a secret key $sk$.
	
	\item \textit{Encaps}($pk$)$\overset{\$}{\rightarrow}$($ct$, $ss$): A encapsulation algorithm that takes the public key $pk$ as input, and outputs a ciphertext $ct$ and a shared secret $ss$.
	
	\item \textit{Decaps}($ct$, $sk$)$\rightarrow$($ss$): A decapsulation algorithm that takes the ciphertext $ct$ and secret key $sk$ as input and outputs the same shared secret $ss$ as algorithm \textit{Encaps} if succeed.
\end{itemize}

\begin{algorithm}[t]
	\caption{$KeyGen$}\label{alg:KeyGen}
	\begin{algorithmic}[1]
		\REQUIRE uniformly random $ (s, seed_{\mathbf{SE}}, z) $
		\STATE $ seed_{\mathbf{A}} \leftarrow \mathrm{SHAKE} (z) $
		\STATE $ \mathbf{A} \leftarrow \textsf{Gen} (seed_{\mathbf{A}}) $
		\STATE $ \mathbf{S}, \mathbf{E} \leftarrow \textsf{Sample} (seed_{\mathbf{SE}}) $
		\STATE $ \mathbf{B} \leftarrow \mathbf{A} \mathbf{S} + \mathbf{E} \mod{q} $
		\STATE $ pkh \leftarrow \mathrm{SHAKE} (seed_{\mathbf{A}} || \mathbf{B}) $
		\STATE $ pk \leftarrow (seed_{\mathbf{A}} || \mathbf{B}), sk \leftarrow (s || seed_{\mathbf{A}} || \mathbf{B}, \mathbf{S}^\mathrm{T}, pkh)$
		\RETURN $ (pk, sk) $
	\end{algorithmic}
\end{algorithm}

\begin{algorithm}[!t]
	\caption{$Encaps$}\label{alg:Encaps}
	\begin{algorithmic}[1]
		\REQUIRE $ pk $, uniformly random $ (u, salt) $
		\STATE $ pkh \leftarrow \mathrm{SHAKE} (pk) $
		\STATE $ seed_{\mathbf{SE}} || k \leftarrow \mathrm{SHAKE} (pkh || u || salt) $
		\STATE $ \mathbf{S}', \mathbf{E}', \mathbf{E}'' \leftarrow \textsf{Sample} (seed_{\mathbf{SE}}) $
		\STATE $ \mathbf{A} \leftarrow \textsf{Gen} (seed_{\mathbf{A}}) $
		\STATE $ \mathbf{B}' \leftarrow \mathbf{S}' \mathbf{A} + \mathbf{E}' \mod{q} $
		\STATE $ \mathbf{C} \leftarrow \mathbf{S}' \mathbf{B} + \mathbf{E}'' + \textsf{Encode} (u) \mod{q} $
		\STATE $ ss \leftarrow \mathrm{SHAKE} (\mathbf{B}' || \mathbf{C} || salt || k) $
		\STATE $ ct \leftarrow (\mathbf{B}' || \mathbf{C}, salt) $
		\RETURN $ (ct,ss) $
	\end{algorithmic}
\end{algorithm}

\begin{algorithm}[!t]
	\caption{$Decaps$}\label{alg:Decaps}
	\begin{algorithmic}[1]
		\REQUIRE $ ct, sk $
		\STATE $ \mathbf{M} \leftarrow \mathbf{C} - \mathbf{B}' \mathbf{S} \mod{q} $
		\STATE $ u' \leftarrow \textsf{Decode} (\mathbf{M}) $
		\STATE $ seed_{\mathbf{SE}}' || k' \leftarrow \mathrm{SHAKE} (pkh || u' || salt) $
		\STATE $ \mathbf{S}', \mathbf{E}', \mathbf{E}'' \leftarrow \textsf{Sample} (seed_{\mathbf{SE}}') $
		\STATE $ \mathbf{A} \leftarrow \textsf{Gen} (seed_{\mathbf{A}}) $
		\STATE $ \mathbf{B}'' \leftarrow \mathbf{S}' \mathbf{A} + \mathbf{E}' \mod{q} $
		\STATE $ \mathbf{C}' \leftarrow \mathbf{S}' \mathbf{B} + \mathbf{E}'' + \textsf{Encode} (u') \mod{q} $
		\STATE $ \bar{k} \leftarrow k' $ if $ (\mathbf{B}' || \mathbf{C}) = (\mathbf{B}'' || \mathbf{C}') $ else $ \bar{k} \leftarrow s $
		\STATE $ ss \leftarrow \mathrm{SHAKE} (\mathbf{B}' || \mathbf{C} || salt || \bar{k}) $
		\RETURN $ ss $
	\end{algorithmic}
\end{algorithm}

Algorithms~\ref{alg:KeyGen}, \ref{alg:Encaps}, and \ref{alg:Decaps} shows the detailed procedures. For clarity, Some steps are described in a simplified manner following the structure of the standard specification.

Key generation first derives seed $seed_{\mathbf{A}}$ using \mbox{SHAKE} and employs it to generate the public matrix $\mathbf{A}$. Independently, the secret matrix $\mathbf{S}$ and the error matrix $\mathbf{E}$ are expanded and sampled from $seed_{\mathbf{SE}}$ using \textsf{Sample}. Next, the LWE matrix multiplication $\mathbf{B} = \mathbf{A}\mathbf{S} + \mathbf{E}$ is computed, which dominates the computational cost of key generation. Finally, a hash value $pkh$ is computed and stored as part of both the public and secret keys.

Encapsulation starts by hashing the public key to obtain $pkh$, which is then used to derive a fresh sampling seed $seed_{\mathbf{SE}}$ (distinct from that used in key generation) together with an intermediate key $k$. Based on $seed_{\mathbf{SE}}$, the ephemeral secret matrix $\mathbf{S}'$ and the error matrices $\mathbf{E}', \mathbf{E}''$ are generated using \textsf{Sample}. The public matrix $\mathbf{A}$ is reconstructed from $seed_{\mathbf{A}}$, and LWE matrix multiplications are performed to generate the ciphertext components, where the computation of $\mathbf{B}' = \mathbf{S}'\mathbf{A} + \mathbf{E}'$ dominates the encapsulation cost. Finally, the shared secret is derived by hashing the ciphertext together with the intermediate key $k$.

Decapsulation first computes the matrix expression $\mathbf{M} = \mathbf{C} - \mathbf{B}'\mathbf{S}$ as part of the decryption process, from which the embedded message $u'$ is recovered using \textsf{Decode}. The recovered message is then used to regenerate the sampling seed and the intermediate key, and the ciphertext is recomputed through a re-encryption process similar to encapsulation. A ciphertext consistency check is performed to ensure correctness under chosen-ciphertext attacks (CCA). Depending on the comparison result, the corresponding key material is selected and hashed together with the ciphertext to produce the final shared secret.

For ease of presentation, $\mathbf{U}=\textsf{Encode}(u)$ and $\mathbf{E}_u=\mathbf{E}''+\textsf{Encode}(u)$ are introduced as auxiliary notations.

The detailed parameters for all 3 security levels are listed in \mbox{Table}~\ref{tab:Parameters}.

\begin{table}[h]
	\centering
	\caption{Parameter Set in FrodoKEM}
	\label{tab:Parameters}
	\begin{tabular}{lcccccccc}
		\toprule
		Parameter Set & $D$ & $q$ & $n$ & $\overline{n}$ & $B$ & $d$ \\
		\toprule
		FrodoKEM-640 & 15 & 32768 & 640 & 8 & 2 & 12 \\
		FrodoKEM-976 & 16 & 65536 & 976 & 8 & 3 & 10 \\
		FrodoKEM-1344 & 16 & 65536 & 1344 & 8 & 4 & 6 \\
		\bottomrule
	\end{tabular}
\end{table}

\begin{figure*}[t]
	\centering
	\includegraphics[width=5.9in]{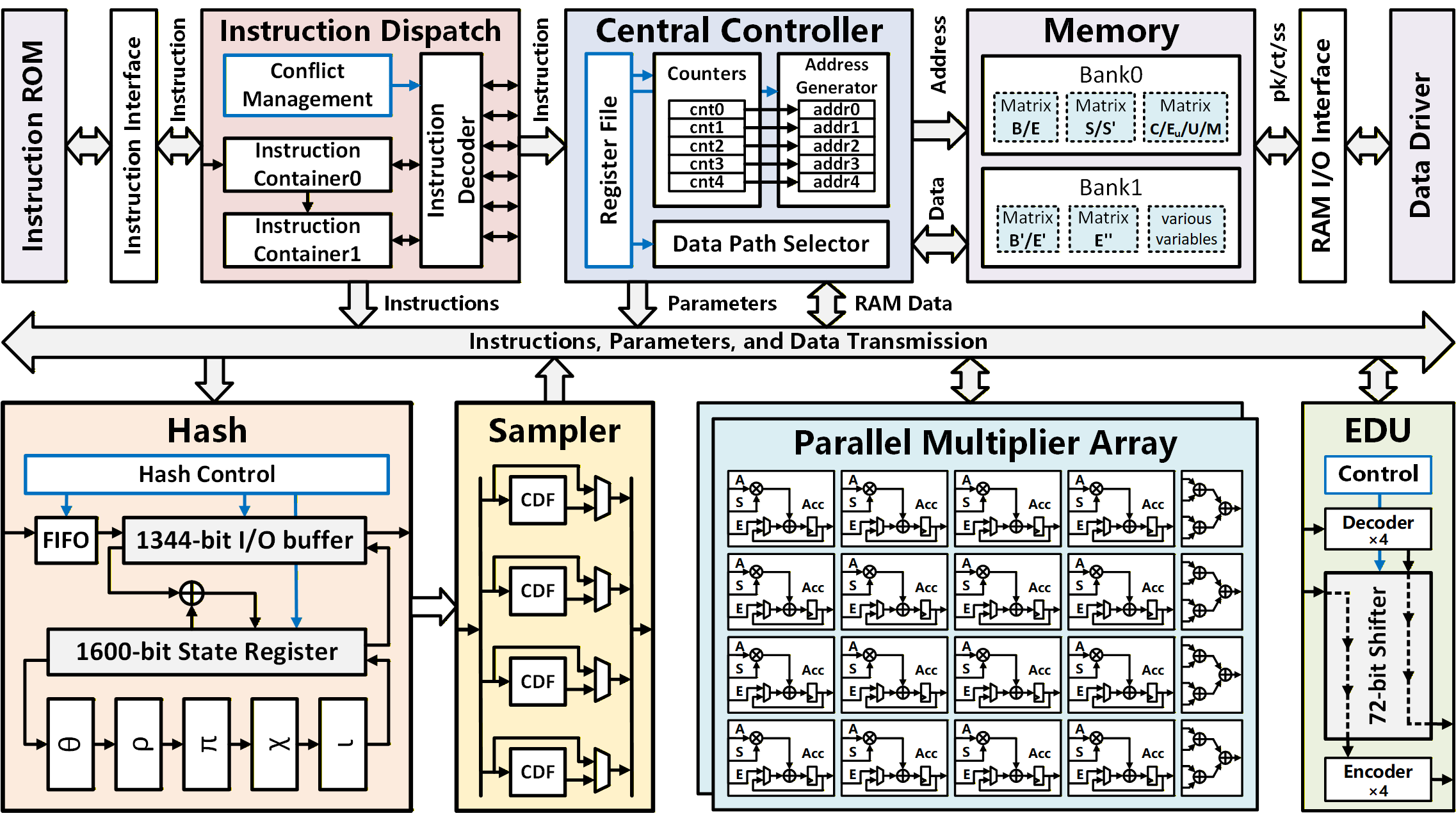}
	\caption{Overall architecture of the FrodoKEM hardware cryptographic processor.}
	\label{fig:overall-architecture}
\end{figure*}

\section{Hardware Design} \label{sec:Hardware}
\subsection{Overall Architecture} \label{subsec:Hardware-Overall}

Fig.~\ref{fig:overall-architecture} illustrates the overall architecture of the proposed FrodoKEM hardware processor.
The processor consists of seven main components: an instruction dispatch unit, a central controller, a hash unit, sampler modules, a parallel multiplier array, an Encode/Decode Unit (EDU), and memory.
The instruction dispatch unit fetches specific instructions and dispatches them to the corresponding functional modules.
The central controller generates memory access addresses and allocates memory ports among the computing units.
Acting as a PRNG, the hash unit produces protocol-required digests (e.g., $pkh$, $ss$) and generates pseudo-random bit streams for subsequent operations.
The sampler modules receive the hash outputs to generate matrix coefficients (e.g., $\mathbf{A}$, $\mathbf{S}$, $\mathbf{E}$, $\mathbf{S}'$, $\mathbf{E}'$, and $\mathbf{E}''$).
To accelerate the most time-consuming matrix computations, a parallel multiplier array is employed.
It supports all matrix operations required across different protocol phases.
The EDU implements the \textsf{Encode} and \textsf{Decode} functions to support protocol-specific encryption and decryption.
The memory stores matrices and other protocol-related variables.
It is partitioned into two banks.
An additional instruction ROM stores instruction sequences for all three security levels and all protocol phases, while a data driver supplies the initial inputs to the processor.

\subsection{Hash and Sampler} \label{subsec:Hardware-Hash}

\begin{figure}[t]
	\centering
	\includegraphics[width=0.95\linewidth]{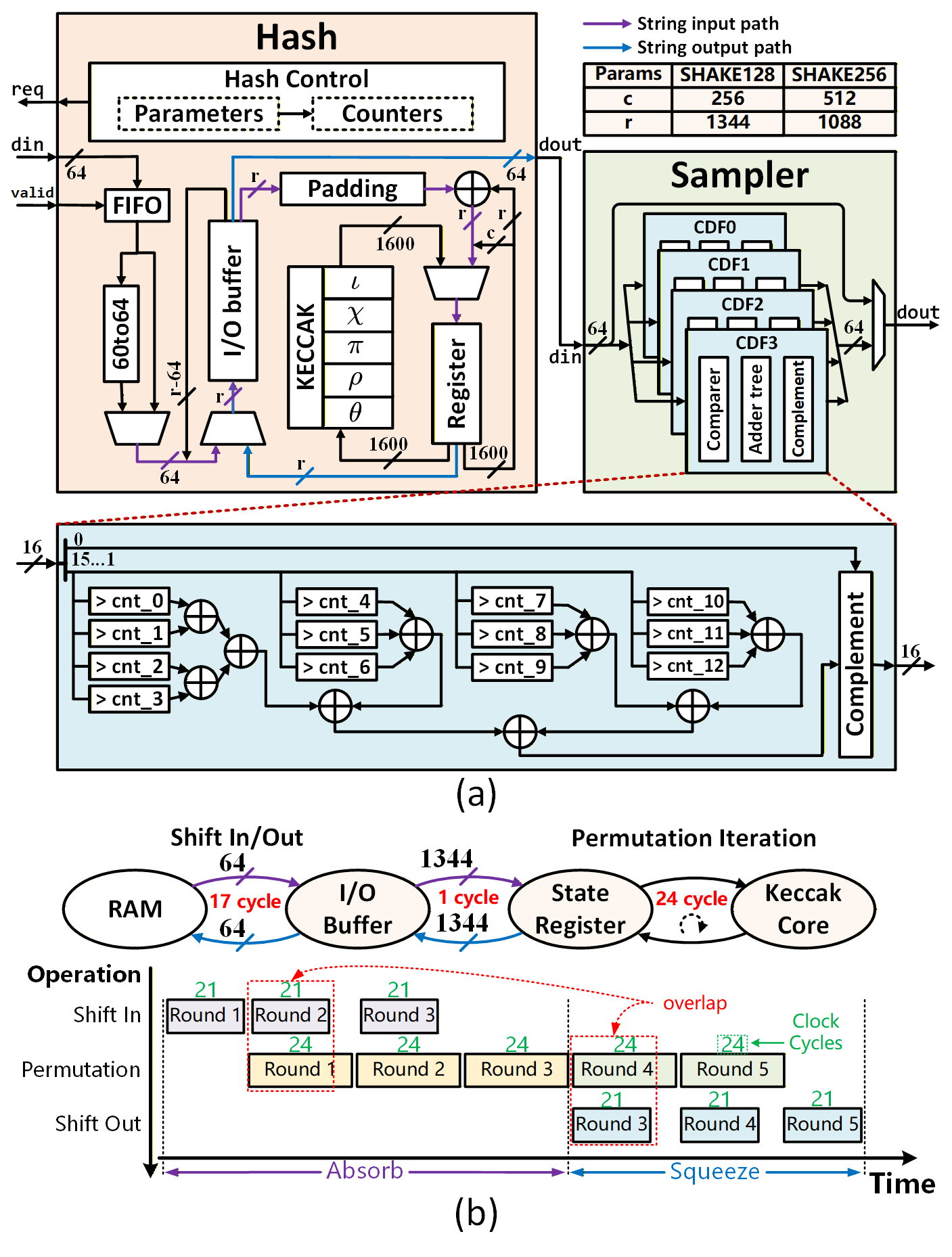}
	\caption{
		(a) Hardware structure of the Keccak-based hash and Gaussian sampler; 
		(b) Dataflow and timing diagram of the buffered hash unit, taking SHAKE128 as an example.}
	\label{fig:hash-sampler}
\end{figure}

The hardware architecture of the hash unit and sampler is illustrated in Fig.~\ref{fig:hash-sampler}(a). 
The hash unit follows the sponge construction with absorb and squeeze phases and supports both SHAKE128 and SHAKE256 via parameter configuration.
It contains a 1600-bit state register, a KECCAK permutation core, and an I/O buffer.
The state register and the KECCAK core form a feedback loop to perform iterative permutations.
Each KECCAK permutation contains 24 rounds and is completed in 24 clock cycles. 
While cascaded KECCAK cores can increase throughput~\cite{Li_2024_High}, they often prolong the critical path and limit the maximum clock frequency. 
For SHAKE128 and SHAKE256, each permutation processes bitstreams of 1344 bits and 1088 bits, respectively.
Accordingly, the I/O buffer is sized to 1344 bits.

The buffer accesses memory through register shifting with a fixed 64-bit bandwidth.
Since 64 is the greatest common divisor of 1088 and 1344, data alignment issues across successive permutations is avoided.
The resulting I/O latency (17 or 21 cycles) closely matches the permutation latency (24 cycles), enabling efficient overlapping between data transfer and permutation.

Fig.~\ref{fig:hash-sampler}(b) illustrates the dataflow and timing for SHAKE128. 
During the absorb phase, the buffer fetches new data while the KECCAK core simultaneously processes the previous data string.
Once a permutation completes, the next data string is instantly injected from the buffer to initiate the subsequent iteration.
The squeeze phase follows the same overlapped scheduling to hide I/O overhead.
This strategy improves the effective throughput by up to $1.84\times$.
Moreover, the same I/O buffer is reused for both absorb and squeeze phases, reducing overall hardware overhead.

Each matrix coefficient is derived from 16-bit data, allowing exactly four coefficients to be generated per cycle.
Accordingly, the sampler integrates four parallel CDF datapaths to match the throughput of the hash unit.
It consumes the hash output and produces matrix coefficients in a constant-time manner to prevent timing side-channel attacks.
Each CDF datapath employs parallel comparators followed by an adder tree to sustain high sampling throughput.

\subsection{Parallel Multiplier Array} \label{subsec:Hardware-Multiplier}

\begin{figure}[t]
	\centering
	\includegraphics[width=0.95\linewidth]{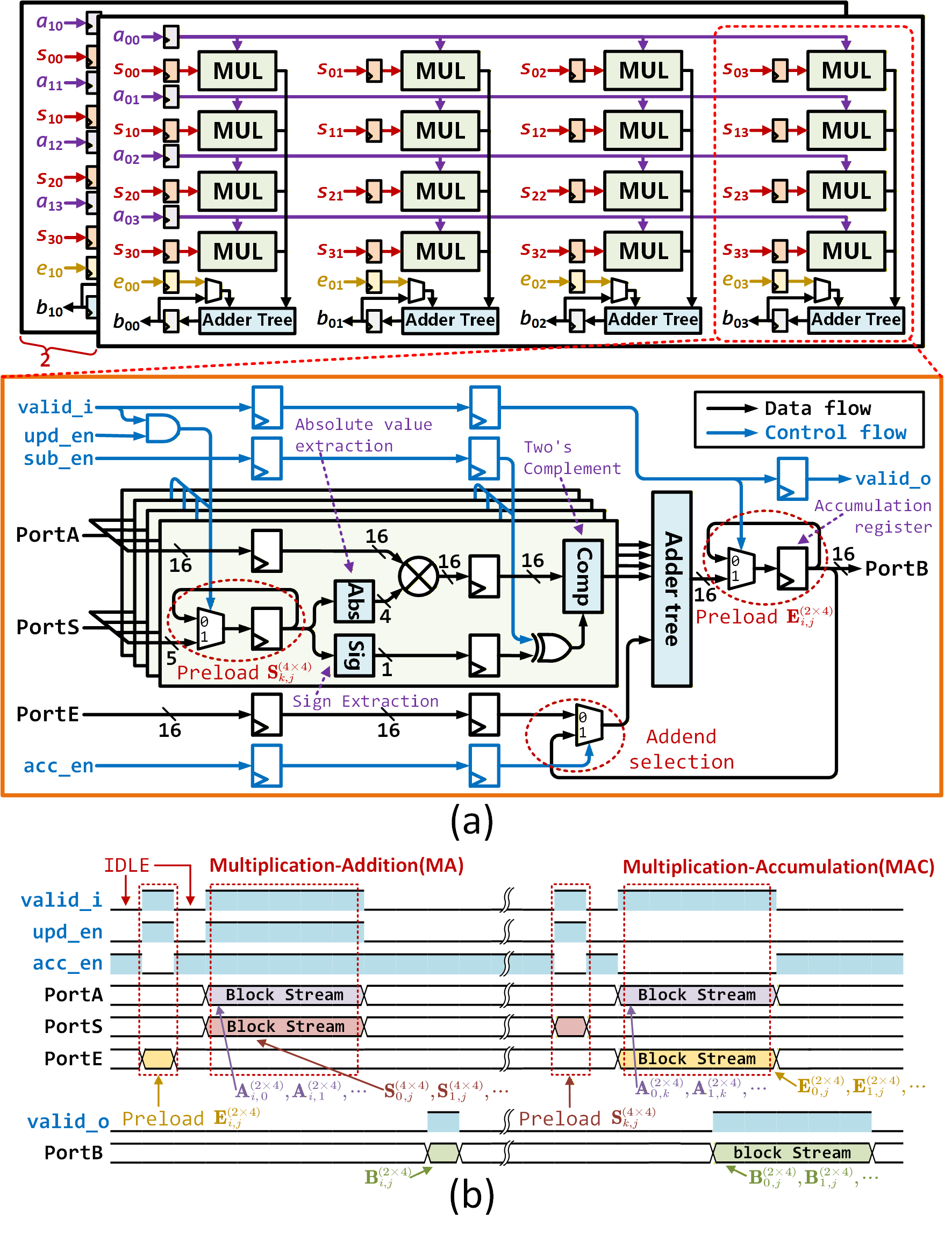}
	\caption{
		(a) Hardware architecture of the multiplier array supporting two computation modes;
		(b) Timing diagram under different computation modes.}
	\label{fig:multiplier-array}
\end{figure}

\begin{figure}[t]
	\centering
	\includegraphics[width=\linewidth]{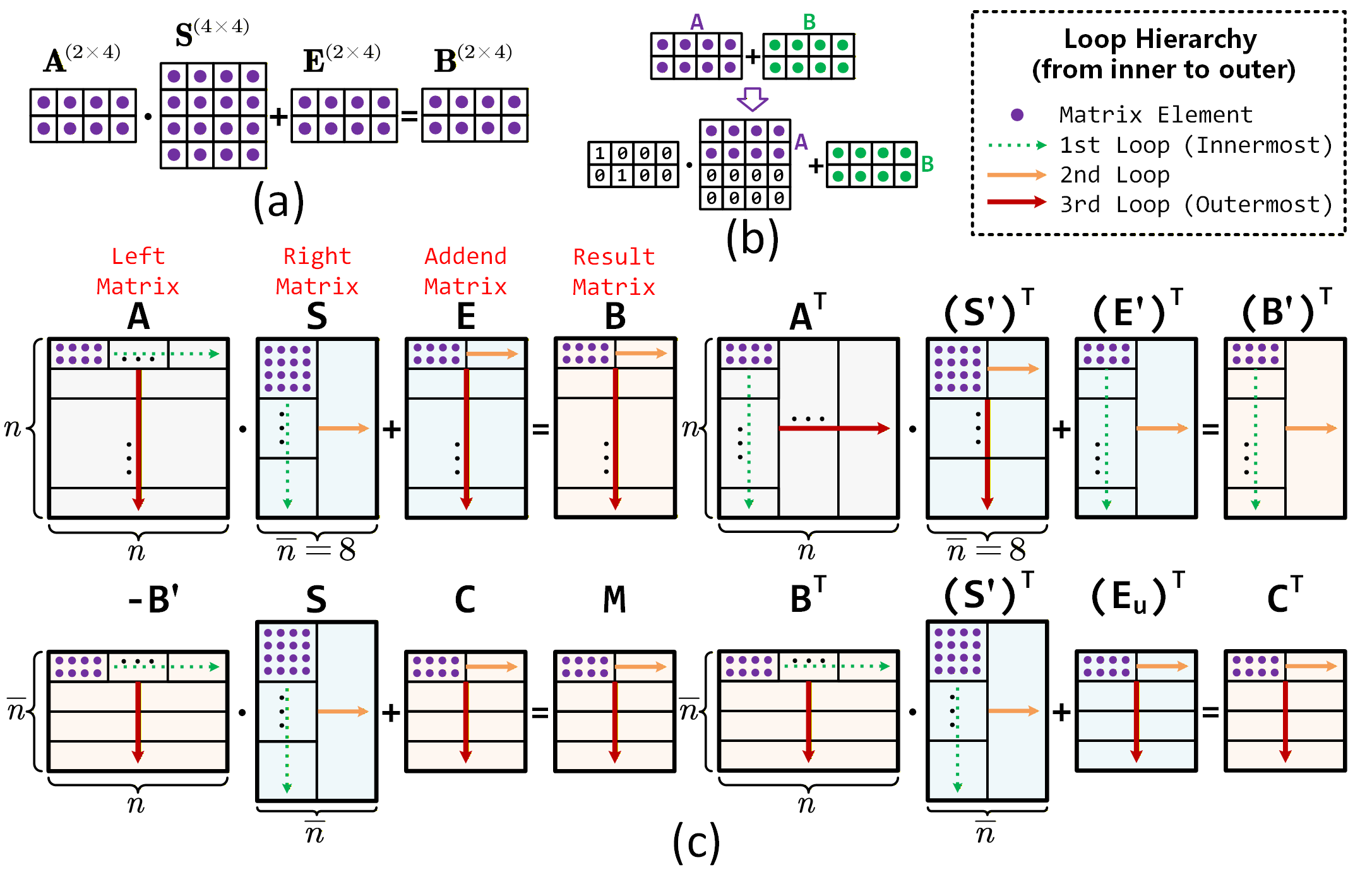}
	\caption{Block partitioning and hierarchical scheduling strategy for matrix multiplication:
		(a) Core block-level computation as the basic form of block-wise processing;
		(b) Block-wise addition implemented by reusing existing hardware resources;
		(c) Block decomposition and loop hierarchy for structured matrix computations.}
	\label{fig:matrix-computation}
\end{figure}

Matrix computations dominate the execution time and account for over 95\% of the total cycles.
This substantial latency makes parallel acceleration essential.
Recent designs have elevated multiplication parallelism to 32~\cite{Duzyol_2025_Can}.
Meanwhile, matrix $\mathbf{A}$ is generated row-wise sequentially.
As a result, different matrix operations exhibit distinct access and scheduling patterns, imposing strong demands on computational flexibility.
To address these challenges, a high-speed and reconfigurable parallel multiplier array is proposed.

As illustrated in Fig.~\ref{fig:multiplier-array}(a), the proposed array consists of 32 multipliers and eight adder trees.
A multi-stage pipeline separates multiplication and accumulation stages, thereby shortening the critical path.
In each cycle, the array performs a block-level operation in the form of
$\mathbf{B}^{(2\times 4)} = \mathbf{A}^{(2\times 4)} \mathbf{S}^{(4\times 4)} + \mathbf{E}^{(2\times 4)}$,
corresponding to the basic block computation in Fig.~\ref{fig:matrix-computation}(a).
The symbols introduced in this subsection are for illustration only and do not refer to specific matrices in the algorithm.
Operands may originate from input ports or internal registers.
Entries of $\mathbf{A}^{(2\times 4)}$, $\mathbf{E}^{(2\times 4)}$, and $\mathbf{B}^{(2\times 4)}$ belong to $\mathbb{Z}_q$ and are represented with 16 bits.
Entries of $\mathbf{S}^{(4\times 4)}$ are signed sampling results with a minimum 5-bit representation.

Each multiplier computes a modular product between a 16-bit operand and a 5-bit signed operand, where modular reduction under $q$ is realized by truncation.
The signed operand is decomposed into a 1-bit sign and a 4-bit absolute value.
The absolute value is multiplied with the 16-bit operand using a $16\times4$ unsigned multiplier, which can be implemented with three adders.
The sign is then recovered according to the sign bit and a subtraction enable signal.
Under modulus $q$, this procedure is mathematically equivalent to direct signed multiplication.
Compared with implementations that map each multiplication to one DSP~\cite{Dang_2019_Implementing,Duzyol_2025_Can}, this design simplifies the arithmetic structure.
Moreover, sign recovery reuses the existing two's-complement logic in the datapath, avoiding dedicated subtraction hardware for $\mathbf{M}=\mathbf{C}-\mathbf{B}'\mathbf{S}$.

The multiplier array is designed to be reconfigurable, supporting two operation modes via mode switching: multiply-accumulate (MAC) and multiply-add (MA).
In MAC mode, a complete block result is obtained within one computation phase as
\begin{equation}
	\mathbf{B}_{i,j}^{(2\times 4)} =
	\mathbf{E}_{i,j}^{(2\times 4)} +
	\sum_{k=0}^{n/4-1}
	\mathbf{A}_{i,k}^{(2\times 4)}
	\mathbf{S}_{k,j}^{(4\times 4)},
	\label{eq:mac_mode}
\end{equation}
where $\mathbf{A}_{i,k}^{(2\times 4)}$, $\mathbf{S}_{k,j}^{(4\times 4)}$, and $\mathbf{E}_{i,j}^{(2\times 4)}$ correspond to blocks from the left matrix, right matrix, and addend matrix in Fig.~\ref{fig:matrix-computation}(c).
Indices $i$ and $j$ denote the block position and remain fixed during one computation phase.
In MA mode, the index $k$ remains fixed, while $i$ is incremented by one in each cycle.
For each $\mathbf{B}_{i,j}^{(2\times 4)}$, one product term is added per cycle, and partial sums are written back immediately.
The final block result is obtained after the last phase.

As shown in Fig.~\ref{fig:multiplier-array}(a), two multiplexers control block preloading, while another multiplexer selects the addend source between the input port and the accumulation register.
Their timing diagram under different modes are illustrated in Fig.~\ref{fig:multiplier-array}(b).
In MAC mode, $\mathbf{E}_{i,j}^{(2\times 4)}$ is preloaded first, followed by streaming $\mathbf{A}_{i,k}^{(2\times 4)}$ and $\mathbf{S}_{k,j}^{(4\times 4)}$ through the input ports to perform accumulation.
In MA mode, $\mathbf{S}_{k,j}^{(4\times 4)}$ is preloaded, then $\mathbf{A}_{i,k}^{(2\times 4)}$ and $\mathbf{E}_{i,j}^{(2\times 4)}$ are streamed, and results $\mathbf{B}_{i,j}^{(2\times 4)}$ are written back to memory.

\subsection{EDU} \label{subsec:Hardware-EDU}

\begin{figure}[t]
	\centering
	\includegraphics[width=\linewidth]{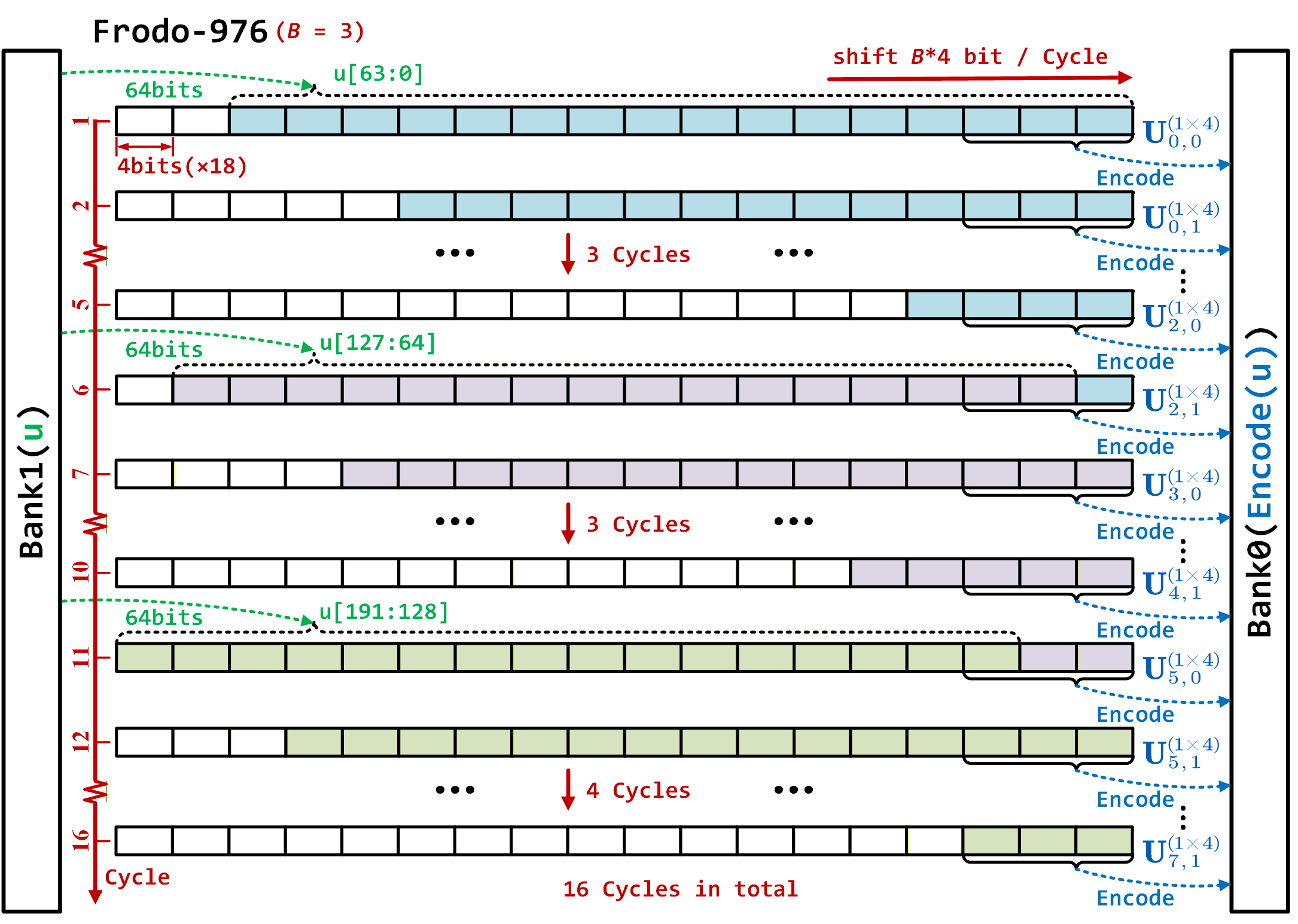} 
	\caption{Dataflow of the shift-register-based data reorganization in the EDU, illustrated by the Encode operation under FrodoKEM-976.}
	\label{fig:encode-decode}
\end{figure}

As discussed in \mbox{Section}~\ref{sec:Preliminary}, each group of $B$ message bits corresponds to a single 16-bit matrix entry.
Since the message granularity does not match the fixed memory access width, shift registers are introduced to buffer and reorganize the data.
The message $u$ is stored in memory Bank1, while the encoded matrix $\mathbf{U}=\textsf{Encode}(u)$ is written to Bank0.
To reduce logic overhead, the EDU shares the same 64-bit memory interface with the hash unit.

Fig.~\ref{fig:encode-decode} illustrates the shift-register-based data reorganization scheme in the EDU, using the Encode operation under FrodoKEM-976 as an example.
A 72-bit shift register is employed and partitioned into 18 segments of 4 bits each.
Compared with the 64-bit memory bandwidth, two extra segments are included.
These two segments tolerate data misalignment across consecutive memory reads, ensuring continuous data delivery without stalling.
During operation, the shift register is shifted by $B$ segments per cycle, producing one $1\times4$ matrix block.
Accordingly, the Encode logic operates with a parallelism of four.
When the remaining valid data in the shift register becomes insufficient, new 64-bit data are fetched from memory and appended to the register, allowing the shifting process to proceed seamlessly.
By iteratively shifting the buffered data, all required Encode outputs are produced in a serialized manner.
The entire shifting process completes within 16 clock cycles.

The same principle applies to other security levels.
Since $4B$ divides 64 under those parameter sets, no additional alignment handling is required, resulting in a simpler dataflow.
The Decode operation follows the reverse dataflow of Encode.
The same hardware structures are reused through appropriate control signals.

\subsection{Instruction Dispatch Unit} \label{subsec:Hardware-Instruction}

\begin{figure}[t]
	\centering
	\includegraphics[width=\linewidth]{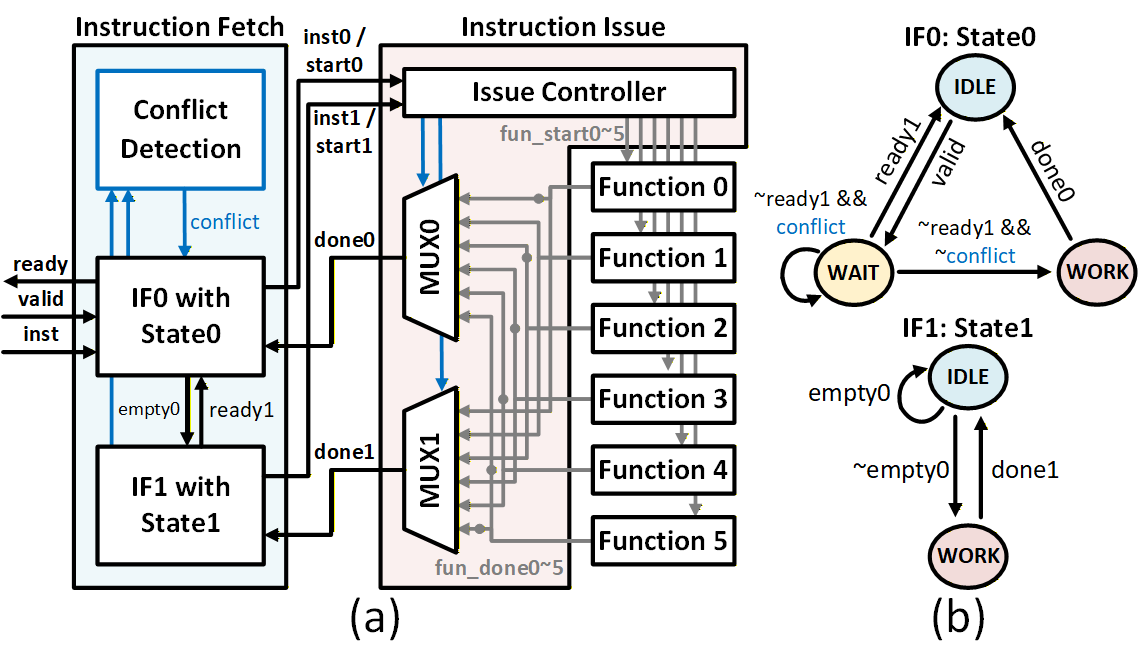} 
	\caption{
		(a) Architectural overview of the instruction dispatch unit, consisting of instruction fetch and issue; 
		(b) State transition diagram for dual-instruction execution;}
	\label{fig:instruction-dispatch}
\end{figure}

The instruction dispatch unit is designed to exploit module-level instruction parallelism in the proposed processor.
Positioned between the instruction fetch interface and the functional modules, it coordinates instruction dispatch across different functional modules.
Since multiple functional modules can operate independently, consecutive instructions without dependencies can be executed in parallel.

As illustrated in Fig.~\ref{fig:instruction-dispatch}(a), the instruction dispatch unit comprises an instruction fetch unit and issue control logic.
The fetch unit utilizes a dual-instruction buffer to temporarily store consecutive instructions.
Conflict detection is then performed based on instruction types and memory access behaviors.
According to the buffer status and conflict detection results, instructions are issued either sequentially or in parallel.
The corresponding state transition behavior follows the scheduling rules shown in Fig.~\ref{fig:instruction-dispatch}(b).
Once issued, instructions are decoded and dispatched to the appropriate functional modules for execution.
This design supports concurrent execution of up to two instructions, enabling effective execution overlap.
By extending the instruction fetch logic, the proposed scheme can be scaled to support overlapped execution of more instructions.

\subsection{Compact Memory Mapping and Organization} \label{subsec:Hardware-Memory}

For large-scale matrix operations, performance is not only determined by the computing parallelism, but also critically constrained by the organization of memory and the efficiency of data access.
Existing designs either rely on substantial memory resources to sustain high parallelism~\cite{Howe_2018_Standard,Duzyol_2025_Can}, or suffer from limited throughput under constrained storage budgets~\cite{Banerjee_2019_Sapphire_ext,Karl_2022_Hardware}.
Therefore, it is necessary to develop a compact memory mapping scheme tailored to block-level matrix computation, together with a flexible dataflow scheduling strategy.

\subsubsection{Matrix Computation Strategy}

Large-scale matrix operations are organized as a sequence of block-level multiplication and addition operations.
Fig.~\ref{fig:matrix-computation}(a) illustrates the core block-level computation pattern, which is directly supported by the parallel multiplier array shown in Section~\ref{subsec:Hardware-Multiplier}.
All matrix operations adhere to this unified block-level pattern, enabling a consistent hardware implementation.
In total, four matrices are computed: $\mathbf{B}$, $\mathbf{B}'$, $\mathbf{M}$, and $\mathbf{C}$.
Their block partitioning and scheduling schemes are summarized in Fig.~\ref{fig:matrix-computation}(c).
For each matrix computation, four operands are involved: the left matrix, the right matrix, the addend matrix, and the result matrix.
As discussed earlier, the right matrices are sampled signed matrices.
For operations involving $\mathbf{S}'$, the computation is reformulated by transposing the expression so that $(\mathbf{S}')^{T}$ appears as the right operand.
This transposition ensures all matrix computations follow a consistent data layout.
Matrices $\mathbf{E}$ and $\mathbf{E}'$ are also generated through sampling.
Since they serve as addends, they are stored in 16-bit format, consistent with $\mathbf{B}$ and $\mathbf{B}'$.
After sampling, sign extension is applied to support modular reduction under modulus $q$.

The overall computation follows a triple-nested loop hierarchy, as shown in Fig.~\ref{fig:matrix-computation}(c).
The innermost loop, indicated by the green dashed arrow, corresponds to one complete computation phase executed by the multiplier array.
The outer two loops advance the block indices along different matrix dimensions, repeatedly invoking the inner-loop computation to cover the entire matrix.
Within each inner-loop computation, the multiplier array operates in either MAC or MA mode.
In MAC mode, the left matrix is accessed row-wise and the right matrix column-wise, with continuous access implemented via a block-wise sliding manner.
The addend and result blocks remain fixed, with one finalized result block produced in each inner-loop iteration.
In MA mode, the left matrix, addend matrix, and result matrix are accessed column-wise, while the right matrix remains fixed.
Here, the computation iterates over the entire result matrix space, progressively updating partial sums until the final result is obtained.
The computations of $\mathbf{B}$, $\mathbf{C}$, and $\mathbf{M}$ employ MAC mode.
Since matrix $\mathbf{A}$ is generated sequentially row-wise, its transposed form $\mathbf{A}^{\mathrm{T}}$ cannot be supplied row-wise for $\mathbf{B}'$ computation.
Thus, MA mode is adopted for $\mathbf{B}'$.

Fully precomputing and storing matrix $\mathbf{A}$ would require 3.45\,MB of storage, equivalent to up to 98 BRAM blocks, imposing significant overhead.
Moreover, since $\mathbf{A}$ is generated sequentially one row at a time, block-level computation demands concurrent access to multiple rows.
Therefore, on-the-fly generation is infeasible; instead, a partially buffered generation strategy is employed.
To minimize latency, $\mathbf{A}$ generation overlaps with matrix computation.
Given the resource intensity of hash functions, only a single hash unit is instantiated.
With a 64-bit output interface, it produces four elements per cycle.
Since the right matrix always has $\bar{n}=8$ columns, each $\mathbf{A}$ element participates in eight multiplications.
To match this production rate, the multiplier array is configured with a parallelism of 32.
This setup sustains computation throughput without underutilizing resources.

During the computation of $\mathbf{C} \leftarrow \mathbf{S}' \mathbf{B} + \mathbf{E}'' + \textsf{Encode}(u)$, the intermediate matrix $\mathbf{E}_u = \mathbf{E}'' + \textsf{Encode}(u)$ is precomputed.
This matrix addition reuses the existing computation structure.
As illustrated in Fig.~\ref{fig:matrix-computation}(b), each operation adds a pair of $2 \times 4$ blocks.
By traversing all block positions, the complete matrix addition is achieved.

\subsubsection{Dataflow Scheduling}

\begin{figure}[t]
	\centering
	\includegraphics[width=0.85\linewidth]{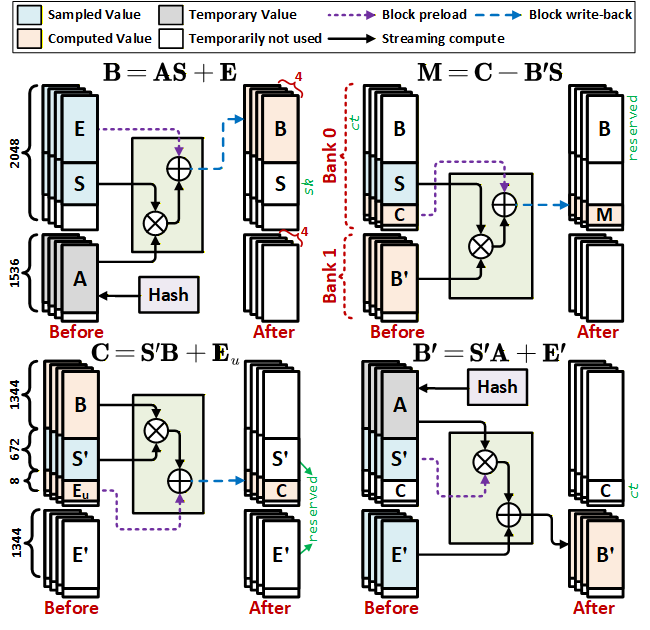}
	\caption{Overall matrix storage scheme and dataflow diagram for matrix computations.}
	\label{fig:memory-dataflow}
\end{figure}

\begin{figure}[t]
\centering
\includegraphics[width=0.9\linewidth]{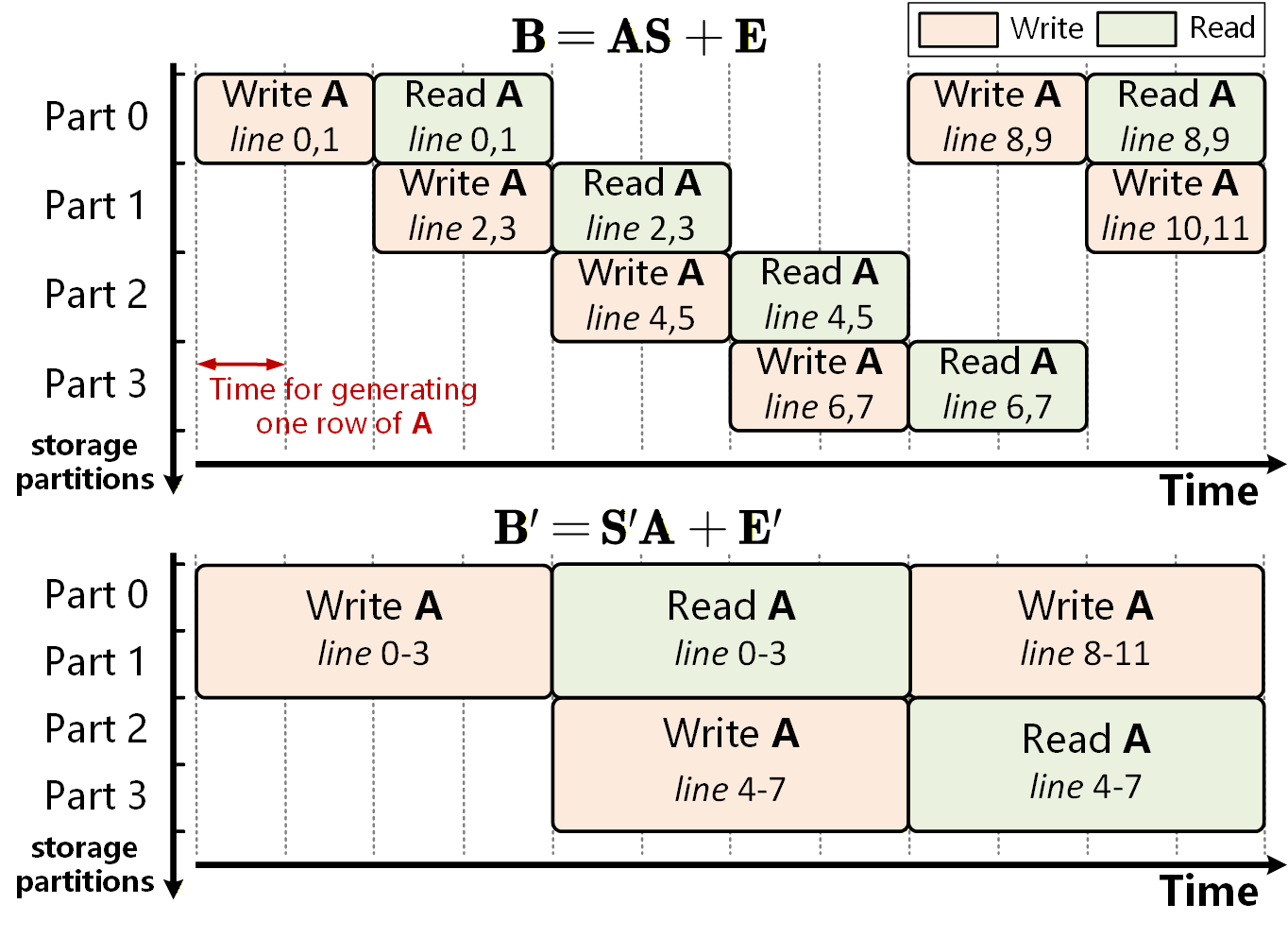}
\caption{ping-pong memory access strategy.}
\label{fig:ping-pong}
\end{figure}

Efficient memory management is essential due to the presence of multiple large matrices, including $\mathbf{A}$, $\mathbf{S}$, $\mathbf{E}$, $\mathbf{B}$, $\mathbf{S}'$, $\mathbf{E}'$, and $\mathbf{B}'$.
As shown in Fig.~\ref{fig:memory-dataflow}, a compact time-multiplexed memory scheduling scheme is employed to minimize BRAM usage.
Specifically, the memory space allocated for $\mathbf{E}$ (denoted as space~$E$) is reused to store $\mathbf{B}$ after its computation, and space~$E'$ is similarly reused for $\mathbf{B}'$.
Matrices $\mathbf{S}$ and $\mathbf{S}'$ share the same memory space (space~$S$), as they are never accessed concurrently.

Matrix $\mathbf{A}$ is generated dynamically and temporarily stored in the reused memory regions. 
These regions are organized into four partitions, each holding two rows of $\mathbf{A}$.
To prevent pipeline stalls during computations of $\mathbf{B}$ and $\mathbf{B}'$, a ping-pong buffering scheme is adopted, as shown in Fig.~\ref{fig:ping-pong}. 
During computation, newly generated rows are written into free partitions, while previously generated rows are simultaneously read for block-wise multiplication.
The read and write roles of the partitions alternate over time.
Since the throughput of the hash unit matches that of the multiplier array, the production and consumption rates of $\mathbf{A}$ are well balanced.

Elements of $\mathbf{S}$ and $\mathbf{S}'$ are stored in 8-bit format, while all other matrices use 16-bit representation.
According to the computation scheme in Fig.~\ref{fig:matrix-computation}(c), all matrices are accessed with a 128-bit bandwidth, which facilitates memory splicing.
The required depths of space~$E$, space~$E'$, and space~$S$ are 1344, 1344, and 672, respectively. 
By concatenating space~$E$ and space~$S$, the total depth approaches 2048, making it suitable for BRAM implementation. 
The overall memory is divided into two banks: Bank~0 stores the spliced spaces and small auxiliary matrices, while Bank~1 holds $\mathbf{B}'$ and other temporary variables.
Compared with separate allocation, this organization reduces BRAM usage by two blocks, resulting in a total of 14 BRAMs.

Additional scheduling optimizations further reduce memory requirements. 
In both \textit{encaps} and \textit{decaps}, the computation order of $\mathbf{C}$ and $\mathbf{B}'$ is swapped. 
This allows the lifespan of $\mathbf{B}$ to terminate earlier, freeing its memory space for storing $\mathbf{B}'$.
At the end of \textit{decaps}, the values $(\mathbf{B}' \| \mathbf{C})$ and $(\mathbf{B}'' \| \mathbf{C}')$ must be compared to verify decryption correctness.
To reduce memory overhead, a hash-packed verification strategy is adopted, inspired by~\cite{Bos_2023_Enabling}.
Instead of storing and directly comparing $(\mathbf{B}' \| \mathbf{C})$ and $(\mathbf{B}'' \| \mathbf{C}')$, the following SHAKE hashes are computed:
\begin{equation}
	\left\{
	\begin{aligned}
		ss_0 &\leftarrow \mathrm{SHAKE}(\mathbf{B}' \| \mathbf{C} \| \text{salt} \| k'),\\
		ss_1 &\leftarrow \mathrm{SHAKE}(\mathbf{B}' \| \mathbf{C} \| \text{salt} \| s),\\
		ss_2 &\leftarrow \mathrm{SHAKE}(\mathbf{B}'' \| \mathbf{C}' \| \text{salt} \| k')
	\end{aligned}
	\right.
	\label{eq:ss}
\end{equation}
Decryption success is verified by comparing $ss_0$ and $ss_2$. 
If they are equal, set $ss = ss_0$; otherwise, set $ss = ss_1$. 
This method reduces memory usage significantly while maintaining security.

\subsubsection{Memory Mapping Scheme}

\begin{figure}[t]
	\centering
	\includegraphics[width=0.95\linewidth]{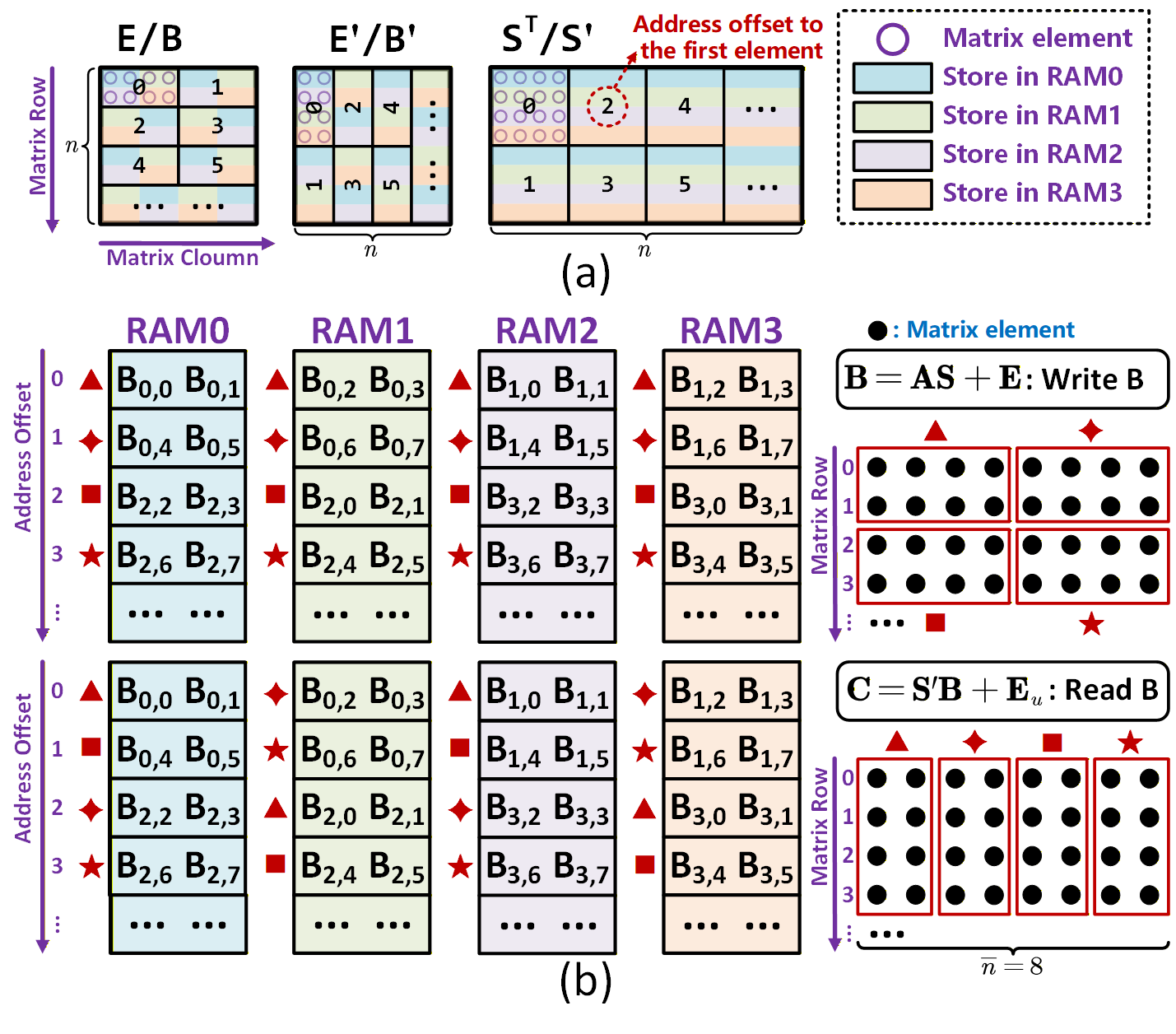}
	\caption{Memory interleaving and access mechanism for matrices: 
		(a) element-to-RAM mapping of matrices; 
		(b) Example of block access scheduling across interleaved RAMs, illustrated using accesses of matrix $\mathbf{B}$ during $\mathbf{B}$ and $\mathbf{C}$ computations.
	}
	\label{fig:address-mapping}
\end{figure}

The storage strategy for matrix elements is determined by their generation order and access patterns. 
To enable flexible memory accesses, both Bank~0 and Bank~1 are physically partitioned along the word width into four RAM blocks, denoted as RAM0--RAM3.
The resulting storage mapping is illustrated in Fig.~\ref{fig:address-mapping}(a), where colors indicate the assigned RAM block for each element, and annotated numbers denote the relative address offsets from the first element in that block.

Each RAM block uses a 32-bit word width, storing either four elements of $\mathbf{S}$/$\mathbf{S}'$ or two elements of $\mathbf{E}$/$\mathbf{E}'$. 
For matrices $\mathbf{S}^T$ and $\mathbf{S}'$, four elements are generated row-wise per cycle and written to a single RAM block. 
During reads, all four RAM blocks are accessed in parallel, retrieving a $4 \times 4$ matrix block per cycle.
For matrices $\mathbf{E}$ and $\mathbf{E}'$, four elements are similarly generated row-wise but distributed across two RAM blocks, with parallel reads yielding a $2 \times 4$ matrix block per cycle.

To enable flexible reshaping of matrix blocks, a memory interleaving scheme is employed for $\mathbf{E}$ and $\mathbf{E}'$. 
Fig.~\ref{fig:address-mapping}(b) shows these access patterns using matrix $\mathbf{B}$ as an example: 
blocks are written in $2 \times 4$ format during $\mathbf{B}$ computation and read in $4 \times 2$ format during $\mathbf{C}$ computation.
Interleaving elements across the four RAM blocks accommodates both patterns without extra data reorganization.
For instance, accessing $\mathbf{B}_{0,0}^{(2 \times 4)}$ sets the relative address to 0 across all RAM blocks.
In contrast, accessing $\mathbf{B}_{0,0}^{(4 \times 2)}$ assigns relative address 0 to RAM0 and RAM2, and 2 to RAM1 and RAM3.
These assignments are indicated by the triangular markers in Fig.~\ref{fig:address-mapping}(b).
All memory addresses are generated by the centralized controller through simple combinational logic.
This flexible and lightweight memory access mechanism enables high throughput.

\begin{figure*}[t]
	\centering
	\includegraphics[width=\linewidth]{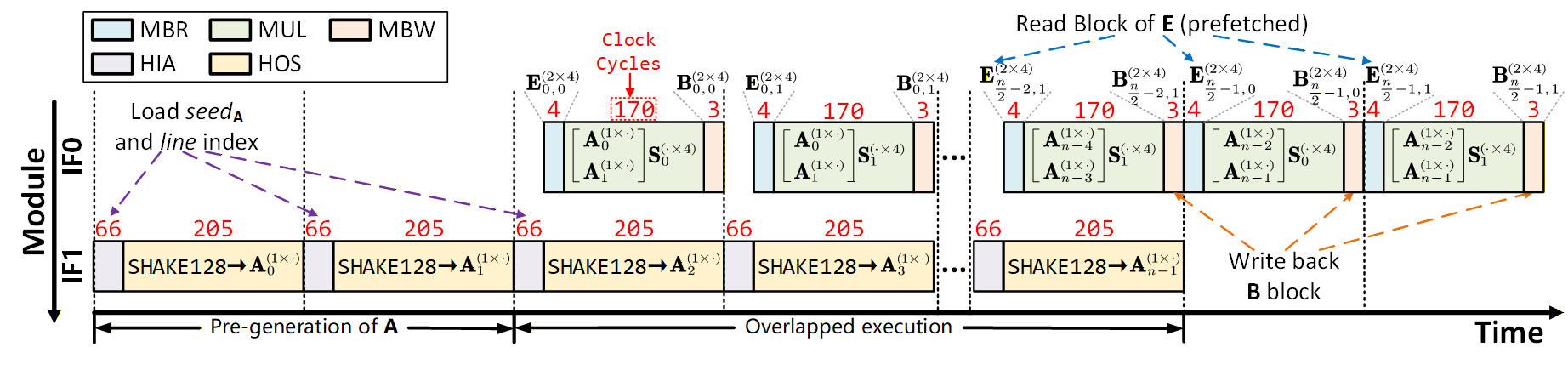} 
	\caption{Timing diagram of instruction-level overlapped execution, illustrated by the computation of $\mathbf{B}$ in FrodoKEM-640.}
	\label{fig:parallel-execution}
\end{figure*}

\subsection{Instruction-Based Overlapped Execution} \label{subsec:Hardware-Overlapped}

\begin{table}[t]
	\centering
	\renewcommand{\arraystretch}{1.0}
	\setlength{\tabcolsep}{6pt}
	\begin{threeparttable}
		\small
		\centering
		\caption{Custom instruction set for FrodoKEM control}
		\label{tab:instructions}
		\begin{tabular}{c c l}
			\toprule
			Category & Instruction & Description \\
			\toprule
			\multirow{2}{*}{\raisebox{-2ex}{\makecell[c]{\emph{Hash} \\ \emph{Control}}}}
			& \texttt{HIA} & Hash Input Absorption \\
			\cmidrule(lr){2-3}
			& \texttt{HOS} & Hash Output Squeezing \\
			\midrule
			\multirow{3}{*}{\raisebox{-3.4ex}{\makecell[c]{\emph{Matrix} \\ \emph{Operations}}}}
			& \texttt{MBR} & Matrix Block Read \\
			\cmidrule(lr){2-3}
			& \texttt{MBW} & Matrix Block Write \\
			\cmidrule(lr){2-3}
			& \texttt{MUL} & Matrix Multiply Operation \\
			\midrule
			\multirow{3}{*}{\raisebox{-3.4ex}{\makecell[c]{\emph{Protocol-Specific} \\ \emph{Operations}}}}
			& \texttt{ENC} & Message Encoding \\
			\cmidrule(lr){2-3}
			& \texttt{DEC} & Message Decoding \\
			\cmidrule(lr){2-3}
			& \texttt{CMP} & Ciphertext Consistency Check \\
			\bottomrule
		\end{tabular}
	\end{threeparttable}
\end{table}

\begin{table}[t]
	\centering
	\renewcommand{\arraystretch}{1}
	\setlength{\tabcolsep}{10.5pt}
	\begin{threeparttable}
		\small
		\centering
		\caption{Instruction Execution Cycle Counting for \textit{KeyGen} and \textit{Decaps} in FrodoKEM-640}
		\label{tab:instruction_cycles}
		\begin{tabular}{c r r r r}
			\toprule
			\multirow{2}{*}{Instruction} & 
			\multicolumn{2}{c}{\textit{KeyGen}} & 
			\multicolumn{2}{c}{\textit{Decaps}} \\
			\cmidrule(lr){2-3} \cmidrule(lr){4-5}
			& Cycles & (\%) & Cycles & (\%) \\
			\toprule
			\midrule[-0.4em]
			\texttt{HIA} & 47072  & 16.5\% & 48750  & 16.9\%  \\
			\texttt{HOS} & 127869 & 44.9\% & 127902 & 44.3\%  \\
			\texttt{MBR} & 1280   &  0.4\% & 1314   & 0.5\%   \\
			\texttt{MBW} & 1280   &  0.4\% & 32     & $<$0.1\% \\
			\texttt{MUL} & 107520 & 37.7\% & 110232 & 38.2\%  \\
			\texttt{ENC} & -      & -      & 23     & $<$0.1\% \\
			\texttt{DEC} & -      & -      & 23     & $<$0.1\% \\
			\texttt{CMP} & -      & -      & 9      & $<$0.1\% \\
			\midrule
			Total        & 285021 & 100\%  & 288285 & 100\%   \\
			\bottomrule
		\end{tabular}
	\end{threeparttable}
\end{table}

\begin{table}[t]
	\centering
	\renewcommand{\arraystretch}{1.0}
	\setlength{\tabcolsep}{4.8pt}
	\begin{threeparttable}
		\small
		\centering
		\caption{Cycle Comparison With and Without Multiple-Instruction Overlapped Execution Across All Security Levels}
		\label{tab:dual_inst_parallel}
		\begin{tabular}{c c c c c}
			\toprule
			\multirow{2}{*}{\makecell[c]{Security \\ Level}} &
			\multirow{2}{*}{\makecell[c]{Protocol \\ Phase}} &
			\multicolumn{2}{c}{Execution Cycles} &
			\multirow{2}{*}{\makecell[c]{Ratio\tnote{a} \\ (\%)}} \\
			\cmidrule{3-4}
			&  & With (kCCs) & Without (kCCs) &  \\
			\toprule
			\multirow{3}{*}{Frodo-640}
			& \textit{KeyGen} & 178.5 &  295.3 & 60.4\% \\
			& \textit{Encaps} & 182.0 &  296.5 & 61.3\% \\
			& \textit{Decaps} & 183.5 &  298.0 & 61.5\% \\
			\midrule
			\multirow{3}{*}{Frodo-976}
			& \textit{KeyGen} & 371.4 &  631.5 & 58.8\% \\
			& \textit{Encaps} & 377.2 &  633.8 & 59.5\% \\
			& \textit{Decaps} & 379.5 &  636.1 & 59.6\% \\
			\midrule
			\multirow{3}{*}{Frodo-1344}
			& \textit{KeyGen} & 656.6 & 1114.4 & 58.9\% \\
			& \textit{Encaps} & 664.4 & 1117.4 & 59.4\% \\
			& \textit{Decaps} & 667.3 & 1120.4 & 59.5\% \\
			\bottomrule
		\end{tabular}
		\begin{tablenotes}
			\footnotesize
			\item[a] The ratio is defined as With/Without and represents the percentage of execution cycles relative to the baseline implementation without dual-instruction parallelism.
		\end{tablenotes}
	\end{threeparttable}
\end{table}

Table~\ref{tab:instructions} summarizes the custom instruction set implemented in the proposed processor.
The instructions are organized into three categories: hash control, matrix operations, and protocol-specific operations.
Specifically, \texttt{HIA} executes the absorption phase of the hash function, while \texttt{HOS} performs the squeezing phase.
Together, they realize a complete hash operation.
The \texttt{MBR} and \texttt{MBW} instructions handle matrix block preloading and write-back, respectively. 
The \texttt{MUL} instruction performs one iteration of the innermost loop computation illustrated in Fig.~\ref{fig:matrix-computation} and outputs results immediately in MA mode.

Table~\ref{tab:instruction_cycles} reports the instruction-level cycle counts for \textit{KeyGen} and \textit{Decaps} in FrodoKEM-640 (the instruction flow of \textit{Encaps} is analogous and therefore omitted).
As shown, hash output squeezing (\texttt{HOS}) and matrix multiplication (\texttt{MUL}) dominate the execution time, together accounting for over 80\% of the total cycles.
Furthermore, the computations of $\mathbf{B}$ and $\mathbf{B}'$ contribute more than 90\% of the overall latency.
These observations indicate that the key target for performance improvement is the overlap between matrix $\mathbf{A}$ generation and subsequent matrix computations, as discussed in Section~\ref{subsec:Hardware-Memory}.

To address this, an instruction-level overlapped execution scheme is proposed.
This scheme is supported by the instruction dispatch unit described in Section~\ref{subsec:Hardware-Instruction}, which enables issuing two independent instructions with overlapped execution.
Note that \texttt{HIA} and \texttt{HOS} require exclusive access to the hash unit and must be executed sequentially.
In contrast, matrix operation instructions do not contend with \texttt{HOS} for hardware resources.
Memory access conflicts are also avoided using the ping-pong buffering strategy illustrated in Fig.~\ref{fig:ping-pong}.

Fig.~\ref{fig:parallel-execution} depicts the timing diagram of instruction-level overlapped execution during the computation of $\mathbf{B} = \mathbf{A}\mathbf{S} + \mathbf{E}$. 
Initially, the first two rows of $\mathbf{A}$ are generated in advance. 
Subsequently, while \texttt{HOS} continues producing the following rows of $\mathbf{A}$, the previously generated rows are simultaneously consumed by \texttt{MUL}. 
The \texttt{MBR} and \texttt{MBW} instructions are also executed in overlap with \texttt{HOS}.
This pipelined pattern repeats according to the ping-pong buffering mechanism.
The computation of $\mathbf{B}' = \mathbf{S}'\mathbf{A} + \mathbf{E}'$ follows the same principle, except that four rows of $\mathbf{A}$ are pre-generated.
In this case, the \texttt{MBW} instruction is not required.

Table~\ref{tab:dual_inst_parallel} compares the total execution cycles with and without the proposed overlapped execution scheme across all FrodoKEM security levels. 
With instruction-level overlapping enabled, the total cycle count is reduced to approximately 58--61\% of the baseline. 
Because the computation of either $\mathbf{B}$ or $\mathbf{B}'$ dominates all protocol phases, the proposed scheme delivers consistent and significant speedups across the entire workflow.

\section{Results and Comparisons} \label{sec:Results}

\subsection{Implementation Results}

\begin{table}[t]
	\centering
	\renewcommand{\arraystretch}{1.0}
	\setlength{\tabcolsep}{4.0pt}
	\begin{threeparttable}
		\small
		\centering
		\caption{Resource Consumption of Core Modules on Artix-7 FPGA\label{tab:Resource}}
		\begin{tabular}{llll}
			\toprule
			Module & LUTs / \% & FFs / \% & BRAMs \\
			\toprule
			Instruction Dispatch & 136  /  1.0\% &   77 /  1.2\% &  0 \\
			Central Controller   & 2915 / 21.7\% &  880 / 14.5\% &  0 \\
			Hash Unit            & 6766 / 50.4\% & 3304 / 54.6\% &  0 \\
			Sampler              & 70   /  0.5\% &   21 /  0.3\% &  0 \\
			Multiplier Array     & 2957 / 22.0\% & 1144 / 18.9\% &  0 \\
			EDU                  & 420  /  3.1\% &  338 /  5.5\% &  0 \\
			Memory               & 159  /  1.1\% &  278 /  4.6\% & 14 \\
			\midrule
			Total                & 13423         &  6042         & 14 \\
			\bottomrule
		\end{tabular}
	\end{threeparttable}
\end{table}

\begin{table}[t]
	\centering
	\renewcommand{\arraystretch}{1.0}
	\setlength{\tabcolsep}{9pt}
	\begin{threeparttable}
		\small
		\centering
		\caption{Execution Cycles of Main Process in FrodoKEM-640\label{tab:Execution}}
		\begin{tabular}{c l l l}
			\toprule
			Phase & Process & Cycles (CCs) & (\%) \\
			\toprule
			\multirow{3}{*}{\textit{KeyGen}}
			& Hash\tnote{a}                                          & 4738   & 2.6\%  \\
			& $\mathbf{B} = \mathbf{A} \mathbf{S} + \mathbf{E}$      & 173823 & 97.3\% \\
			\cmidrule(lr){2-4}
			& Total                                                  & 178561 & 100\%  \\
			\midrule
			\multirow{6}{*}{\textit{Decaps}}
			& Hash\tnote{a}                                          & 6355   & 3.4\%  \\
			& $\mathbf{B}' = \mathbf{S}' \mathbf{A} + \mathbf{E}'$   & 174166 & 94.8\% \\
			& $\mathbf{M} = \mathbf{C} - \mathbf{B}' \mathbf{S}$     & 1488   & 0.8\%  \\
			& $\mathbf{C} = \mathbf{S}' \mathbf{B} + \mathbf{E}_u$  & 1488   & 0.8\%  \\
			& Others                                                 & 101    & $<$0.1\% \\
			\cmidrule(lr){2-4}
			& Total                                                  & 183598 & 100\%  \\
			\bottomrule
		\end{tabular}
		\begin{tablenotes}
			\footnotesize
			\item[a] The hash cycles here exclude the generation of the public matrix $\mathbf{A}$.
		\end{tablenotes}
	\end{threeparttable}
\end{table}

\begin{table}[t]
	\centering
	\renewcommand{\arraystretch}{1.0}
	\setlength{\tabcolsep}{4.5pt}
	\begin{threeparttable}
		\small
		\centering
		\caption{Latency of Processor on Artix-7/Ultrascale+ FPGA\label{tab:Latency}}
		\begin{tabular}{cccc}
			\toprule
			\multirow{2}{*}{$n$} & \multirow{2}{*}{\tabincell{c}{\textit{KeyGen} \\ (kCCs/ms\tnote{a} /ms\tnote{b} )}} & \multirow{2}{*}{\tabincell{c}{\textit{Encaps} \\ (kCCs/ms\tnote{a} /ms\tnote{b} )}} & \multirow{2}{*}{\tabincell{c}{\textit{Decaps} \\ (kCCs/ms\tnote{a} /ms\tnote{b} )}} \\
			&  &  \\
			\toprule
			640  & 178.5/0.859/0.356 & 182.0/0.876/0.363 & 183.6/0.883/0.366 \\
			976  & 371.4/1.788/0.741 & 377.3/1.815/0.753 & 379.5/1.826/0.757 \\
			1344 & 656.6/3.160/1.310 & 664.4/3.197/1.326 & 667.3/3.212/1.332 \\
			\bottomrule
		\end{tabular}
		\begin{tablenotes}
			\footnotesize
			\item[a] Result for Artix-7 FPGA, with a maximum frequency of 207 MHz.
			\item[b] Result for Ultrascale+ FPGA, with a maximum frequency of 501 MHz.
		\end{tablenotes}
	\end{threeparttable}
\end{table}

The proposed FrodoKEM crypto-processor is implemented in Verilog HDL and verified through Modelsim.
It is implemented on Xilinx Artix-7 and UltraScale+ FPGA platforms to evaluate both resource efficiency and performance.
The implementation provides full hardware support for all FrodoKEM security levels and protocol phases.
On the Artix-7 FPGA, the processor consumes 13,423 LUTs, 6,042 FFs, and 14 BRAMs, achieving the fastest reported execution latency among existing hardware implementations.
Detailed resource breakdown and performance results are presented in Tables~\ref{tab:Resource}, \ref{tab:Execution}, and \ref{tab:Latency}.

Table~\ref{tab:Resource} reports the resource consumption of each core module on the Artix-7 FPGA. 
The hash unit and the parallel multiplier array dominate the logic utilization, accounting for 50.4\% and 22.0\% of LUT usage, respectively. 
Simplifying the multiplication logic allows the design to avoid resource-intensive DSP blocks. 
In addition, the configurable design of the multiplier array and the reuse of input/output buffers in the hash unit further reduce resource consumption.
The instruction dispatch unit consumes only about 1\% of total LUTs due to its lightweight design.
Moreover, a compact memory scheduling strategy shortens the lifetime of intermediate matrices, allowing all storage to be implemented using only 14 BRAMs.

The execution-cycle analysis is summarized in Table~\ref{tab:Execution} and Table~\ref{tab:Latency}.
As shown in Table~\ref{tab:Execution}, the computations of $\mathbf{B}$ and $\mathbf{B}'$ dominate the execution time in both \textit{KeyGen} and \textit{Decaps}, accounting for over 94\% of the total cycles.
This clearly identifies large-scale matrix multiplication as the primary performance bottleneck.
To address this, the proposed architecture overlaps public matrix $\mathbf{A}$ generation with matrix multiplication. 
This is realized through lightweight instruction-level execution overlap, resulting in a $1.65\times$ speedup.
Hash operations are further accelerated by overlapping data transfer and KECCAK permutation, effectively hiding I/O latency and improving throughput by up to $1.84\times$. 
In addition, matrix multiplication is accelerated by a multi-stage pipelined multiplier array, which shortens the critical path and enables higher operating frequencies.
Table~\ref{tab:Latency} summarizes the overall latency across all FrodoKEM security levels and protocol phases.
On the Artix-7 FPGA, the processor achieves a maximum frequency of 207~MHz, completing \textit{KeyGen}/\textit{Encaps}/\textit{Decaps} in 0.86~ms/0.88~ms/0.88~ms for the FrodoKEM-640 security level.

\subsection{Comparison with Related Works}

\begin{table*}[t]
	\begin{spacing}{1.1}
		\setlength{\tabcolsep}{5.4pt}
		\begin{threeparttable}
			
			\small
			\caption{Comparison of Performance and Resource\label{tab:Comparison}}
			\centering
			\begin{tabular}{cccccccccccc}
				\toprule 
				\multirow{2}{*}{\tabincell{c}{Work \\ (Platform)}} & \multirow{2}{*}{Method} & \multirow{2}{*}{\tabincell{c}{Security \\ Level}} & \multicolumn{5}{c}{Area} & \multirow{2}{*}{\tabincell{c}{Frequency \\ (MHz)}} & \multicolumn{3}{c}{Time(ms)/ATP\tnote{b}} \\ 
				
				\cline{4-8} \cline{10-12}
				& & & LUT & FF & DSP & BRAM & Slice\tnote{a} & & $KeyGen$ & $Encaps$ & $Decaps$ \\
				
				\toprule
				\multirow{3}{*}{\tabincell{c}{\textbf{This Work} \\ (Artix-7)}} & 
				\multirow{3}{*}{HW} & 
				\multirow{3}{*}{\tabincell{c}{Frodo640 \\ Frodo976 \\ Frodo1344}} & 
				\multirow{3}{*}{13423} & 
				\multirow{3}{*}{6042} & 
				\multirow{3}{*}{0} & 
				\multirow{3}{*}{14} & 
				\multirow{3}{*}{6609} & 
				\multirow{3}{*}{207} & 
				\multirow{3}{*}{\tabincell{c}{\textbf{0.859/0.56} \\ \textbf{1.788/1.18} \\ \textbf{3.160/2.08}}} & 
				\multirow{3}{*}{\tabincell{c}{\textbf{0.876/0.57} \\ \textbf{1.815/1.20} \\ \textbf{3.197/2.11}}} & 
				\multirow{3}{*}{\tabincell{c}{\textbf{0.883/0.58} \\ \textbf{1.826/1.20} \\ \textbf{3.212/2.12}}} \\
				&  &  &  &  &  &  &  &  &  &  &  \\
				&  &  &  &  &  &  &  &  &  &  &  \\
				
				\midrule 
				
				\multirow{6}{*}{\tabincell{c}{\cite{Howe_2018_Standard}\tnote{c} \\ (Artix-7)}} & 
				\multirow{6}{*}{HW} & 
				\multirow{6}{*}{\tabincell{c}{Frodo640\tnote{d} \\ \ \\ \  \\ Frodo976\tnote{d}}} & 
				\multirow{6}{*}{\tabincell{c}{3771 \\ 6745 \\ 7220 \\ 7139 \\ 7209 \\ 7773}} & 
				\multirow{6}{*}{\tabincell{c}{1800 \\ 3528 \\ 3549 \\ 1800 \\ 3537 \\ 3559}} & 
				\multirow{6}{*}{\tabincell{c}{1 \\ 1 \\ 1 \\ 1 \\ 1 \\ 1}} & 
				\multirow{6}{*}{\tabincell{c}{6 \\ 11 \\ 16 \\ 8 \\ 16 \\ 24}} & 
				\multirow{6}{*}{\tabincell{c}{2439 \\ 4345 \\ 5625 \\ 3746 \\ 5623 \\ 7623}} & 
				\multirow{6}{*}{\tabincell{c}{167 \\ 167 \\ 162 \\ 167 \\ 167 \\ 162}} & 
				\multirow{6}{*}{\tabincell{c}{19.62/4.7 \\ - \\ - \\ 45.63/17 \\ - \\ -}} & 
				\multirow{6}{*}{\tabincell{c}{- \\ 19.86/8.6 \\ - \\ - \\ 46.00/25.8 \\ -}} & 
				\multirow{6}{*}{\tabincell{c}{- \\ - \\ 20.73/11.6 \\ - \\ - \\ 47.81/36.4}} \\
				&  &  &  &  &  &  &  &  &  &  &  \\
				&  &  &  &  &  &  &  &  &  &  &  \\
				\cline{3-12}
				&  &  &  &  &  &  &  &  &  &  &  \\
				&  &  &  &  &  &  &  &  &  &  &  \\
				&  &  &  &  &  &  &  &  &  &  &  \\
				
				\midrule 

				\multirow{3}{*}{\tabincell{c}{\cite{Duzyol_2025_Can} \\ (Artix-7)}} & 
				\multirow{3}{*}{HW} & 
				\multirow{3}{*}{\tabincell{c}{Frodo640\tnote{d}}} & 
				\multirow{3}{*}{\tabincell{c}{12246 \\ 15908 \\ 17066}} & 
				\multirow{3}{*}{\tabincell{c}{8859 \\ 12111 \\ 12868}} & 
				\multirow{3}{*}{\tabincell{c}{24 \\ 32 \\ 32}} & 
				\multirow{3}{*}{\tabincell{c}{18 \\ 20 \\ 20}} & 
				\multirow{3}{*}{\tabincell{c}{9702 \\ 11901 \\ 12191}} & 
				\multirow{3}{*}{\tabincell{c}{147 \\ 136.98 \\ 133.15}} & 
				\multirow{3}{*}{\tabincell{c}{1.02/0.9 \\ - \\ - }} & 
				\multirow{3}{*}{\tabincell{c}{- \\ 0.92/1.1 \\ - }} & 
				\multirow{3}{*}{\tabincell{c}{- \\ - \\ 0.95/1.1 }} \\
				&  &  &  &  &  &  &  &  &  &  &  \\
				&  &  &  &  &  &  &  &  &  &  &  \\
				
				\midrule 
				
				\multirow{3}{*}{\tabincell{c}{\cite{Banerjee_2019_Sapphire_ext} \\ (Artix-7)}} & 
				\multirow{3}{*}{HW/SW} & 
				\multirow{3}{*}{\tabincell{c}{Frodo640 \\ Frodo976 \\ Frodo1344}} & 
				\multirow{3}{*}{\tabincell{c}{14975}} & 
				\multirow{3}{*}{\tabincell{c}{2539}} & 
				\multirow{3}{*}{\tabincell{c}{11}} &
				\multirow{3}{*}{\tabincell{c}{14}} & 
				\multirow{3}{*}{\tabincell{c}{8123}} & 
				\multirow{3}{*}{\tabincell{c}{25}} & 
				\multirow{3}{*}{\tabincell{c}{\ 458/\ 372 \\ 1040/\ 845 \\ 2719/2209}} & 
				\multirow{3}{*}{\tabincell{c}{\ 464/\ 377 \\ 1189/\ 966 \\ 2860/2323}} & 
				\multirow{3}{*}{\tabincell{c}{\ 481/\ 391 \\ 1216/\ 988 \\ 2901/2356}} \\
				&  &  &  &  &  &  &  &  &  &  &  \\
				&  &  &  &  &  &  &  &  &  &  &  \\ 
				
				\midrule 
				\multirow{3}{*}{\tabincell{c}{\textbf{This Work} \\ (UltraScale+)}} & 
				\multirow{3}{*}{HW} & 
				\multirow{3}{*}{\tabincell{c}{Frodo640 \\ Frodo976 \\ Frodo1344}} & 
				\multirow{3}{*}{13271} & 
				\multirow{3}{*}{6067} & 
				\multirow{3}{*}{0} & 
				\multirow{3}{*}{14} & 
				\multirow{3}{*}{3285} & 
				\multirow{3}{*}{501} & 
				\multirow{3}{*}{\tabincell{c}{\textbf{0.356/0.11} \\ \textbf{0.741/0.24} \\ \textbf{1.310/0.43}}} & 
				\multirow{3}{*}{\tabincell{c}{\textbf{0.363/0.11} \\ \textbf{0.753/0.24} \\ \textbf{1.326/0.43}}} & 
				\multirow{3}{*}{\tabincell{c}{\textbf{0.366/0.11} \\ \textbf{0.757/0.24} \\ \textbf{1.332/0.43}}} \\
				&  &  &  &  &  &  &  &  &  &  &  \\
				&  &  &  &  &  &  &  &  &  &  &  \\
				
				\midrule 
				
				\multirow{3}{*}{\tabincell{c}{\cite{Dang_2019_Implementing}\tnote{e} \\ (UltraScale+)}} & 
				\multirow{3}{*}{HW/SW} & 
				\multirow{3}{*}{\tabincell{c}{Frodo640 \\ Frodo976 \\ Frodo1344}} & 
				\multirow{3}{*}{\tabincell{c}{7213 \\ 7087 \\ 7015}} & 
				\multirow{3}{*}{\tabincell{c}{6647 \\ 6693 \\ 6610}} & 
				\multirow{3}{*}{\tabincell{c}{32 \\ 32 \\ 32}} &
				\multirow{3}{*}{\tabincell{c}{13.5 \\ 17 \\ 17.5}} & 
				\multirow{3}{*}{\tabincell{c}{4108 \\ 4499 \\ 4548}} & 
				\multirow{3}{*}{\tabincell{c}{402 \\ 402 \\ 417}} & 
				\multirow{3}{*}{\tabincell{c}{- \\ - \\ -}} & 
				\multirow{3}{*}{\tabincell{c}{1.223/0.50 \\ 1.642/0.73 \\ 2.186/0.99}} & 
				\multirow{3}{*}{\tabincell{c}{1.319/0.54 \\ 1.866/0.83 \\ 3.120/1.41}} \\
				&  &  &  &  &  &  &  &  &  &  &  \\
				&  &  &  &  &  &  &  &  &  &  &  \\
				
				\midrule 
				
				\multirow{3}{*}{\tabincell{c}{\cite{Karl_2022_Hardware} \\ (UltraScale+)}} & 
				\multirow{3}{*}{HW/SW} & 
				\multirow{3}{*}{\tabincell{c}{Frodo640 \\ Frodo976 \\ Frodo1344}} & 
				\multirow{3}{*}{5590} & 
				\multirow{3}{*}{1118} & 
				\multirow{3}{*}{2} &
				\multirow{3}{*}{0} & 
				\multirow{3}{*}{801} & 
				\multirow{3}{*}{100} & 
				\multirow{3}{*}{\tabincell{c}{234/18 \\ 477/38 \\ 854/68}} & 
				\multirow{3}{*}{\tabincell{c}{254/20 \\ 517/41 \\ 924/74}} & 
				\multirow{3}{*}{\tabincell{c}{253/20 \\ 514/41 \\ 920/73}} \\
				&  &  &  &  &  &  &  &  &  &  &  \\
				&  &  &  &  &  &  &  &  &  &  &  \\
				
				\bottomrule 
			\end{tabular}
			\begin{tablenotes}
				\footnotesize
				\item[a] Equivalent Slice is calculated by the formula LUTs/4 + DSPs$\times$102.4 + BRAMs$\times$232.4 for 7 series FPGA and LUTs/8 + DSPs$\times$51.2 + BRAMs$\times$116.2 for Ultrascale+ FPGA\cite{Huang_2025_Lightweight}.
				\item[b] ATP uses Slices $\times$ time for equivalent calculation.
				\item[c] The implementation is based on the NIST Round 1 submission, which supports only two security levels and features a slightly different algorithmic flow compared to the current specification.
				\item[d] Different protocol phases, namely \textit{KeyGen}, \textit{Encaps}, and \textit{Decaps}, are implemented using separate hardware instances.
				The resource usage of these instances is reported in three rows in this order.
				\item[e] \textit{KeyGen} is not implemented, and separate implementations are adopted for different security levels.
			\end{tablenotes}
		\end{threeparttable}
	\end{spacing}
\end{table*}

Table~\ref{tab:Comparison} presents a detailed comparison between the proposed design and existing FPGA implementations, including resource usage, operating frequency, execution time, and ATP for \textit{KeyGen}, \textit{Encaps}, and \textit{Decaps}.

Early FPGA implementations by Howe \emph{et al.}~\cite{Howe_2018_Standard} focus on minimizing logic and memory footprints on Artix-7 devices. 
Their designs instantiate separate hardware modules for different protocol phases and security levels. 
Only a single DSP is employed, which severely limits parallelism in matrix computations.
In addition, with large intermediate data being stored, BRAM consumption increases significantly at higher security levels (e.g., 24 BRAMs for \textit{Decaps} at Frodo-976).
By increasing architectural parallelism, our design runs $25.5\times/25.3\times/26.2\times$ faster in \textit{KeyGen}/\textit{Encaps}/\textit{Decaps} at Frodo-976.
Correspondingly, the ATP is improved by $14.5\times/21.6\times/30.2\times$, demonstrating substantially higher hardware efficiency.

A more recent standard-compliant implementation is reported by Düzyol \emph{et al.}~\cite{Duzyol_2025_Can}. 
High throughput on Artix-7 is achieved by increasing DSP and BRAM usage.  
Two distinct architectures are adopted for $\mathbf{A}\!\cdot\!\mathbf{S}$ and $\mathbf{S}'\!\cdot\!\mathbf{A}$ multiplications.
For $\mathbf{A}\!\cdot\!\mathbf{S}$, a dedicated architecture with a parallelism of 24 is employed.
For $\mathbf{S}'\!\cdot\!\mathbf{A}$, parallelism is increased to 32 by instantiating two hash units, further raising resource consumption.
In contrast, our design employs a single hash unit together with a 32-way parallel multiplier array, reducing execution time by 6\%--16\%. 
Moreover, their memory scheduling does not promptly release intermediate data, leading to significant storage overhead (e.g., 20 BRAMs for \textit{Decaps} at Frodo-640).
Our compact memory scheduling shortens the lifespan of intermediate variables and lowers overall memory demand, yielding $1.75\times/1.91\times/2.00\times$ ATP improvements for \mbox{\textit{KeyGen}}/\textit{Encaps}/\textit{Decaps}, respectively.

Dang \emph{et al.}~\cite{Dang_2019_Implementing} accelerate matrix generation and multiplication through a software/hardware co-design on UltraScale+ platforms, achieving millisecond-level latency. 
However, each security level requires a dedicated hardware implementation, and \textit{KeyGen} is not supported, which limits practical deployment.
Control and scheduling rely on an external high-speed processor, and additional AXI DMA and FIFO components are required for coordination.
Compared with this approach, our design achieves up to $3.6\times$ lower execution time and improves ATP by up to $4.5\times$.

Other software/hardware co-designs, such as Sapphire~\cite{Banerjee_2019_Sapphire_ext} and the RISC-V-based architecture in~\cite{Karl_2022_Hardware}, support all security levels and protocol phases. 
They accelerate execution by offloading compute-intensive kernels to hardware accelerators. 
Nevertheless, overall execution time remains on the order of hundreds of milliseconds due to limited parallelism in large matrix operations. 
Moreover, their hardware is not optimized for high operating frequencies. 
In contrast, our fully hardware implementation attains a high clock frequency through deep pipelining and exploits parallel hardware structures, achieving speedups of several hundred times.

\section{Conclusion} \label{sec:Conclusion}

This paper presents a high-speed and efficient crypto-processor for FrodoKEM with full support for all security levels and protocol phases.
A multi-instruction parallel execution scheme leverages instruction-level concurrency to minimize execution latency.
Computationally intensive matrix operations are accelerated by a high-speed, reconfigurable parallel multiplier array.
Furthermore, a compact memory scheduling strategy is employed to reduce overall storage consumption while sustaining high throughput.
Consequently, the proposed processor utilizes merely 6609 equivalent slices on an Artix-7 FPGA, attaining 1.75$\times$-2.00$\times$ ATP enhancements and up to 16\% reductions in execution time compared with \mbox{state-of-the-art} implementations.

Future work will focus on optimizing the design of specific instructions and enabling tighter integration with \mbox{RISC-V} processors.


\vspace{11pt}

\begin{IEEEbiography}[{\includegraphics[width=1in,height=1.25in,clip,keepaspectratio]{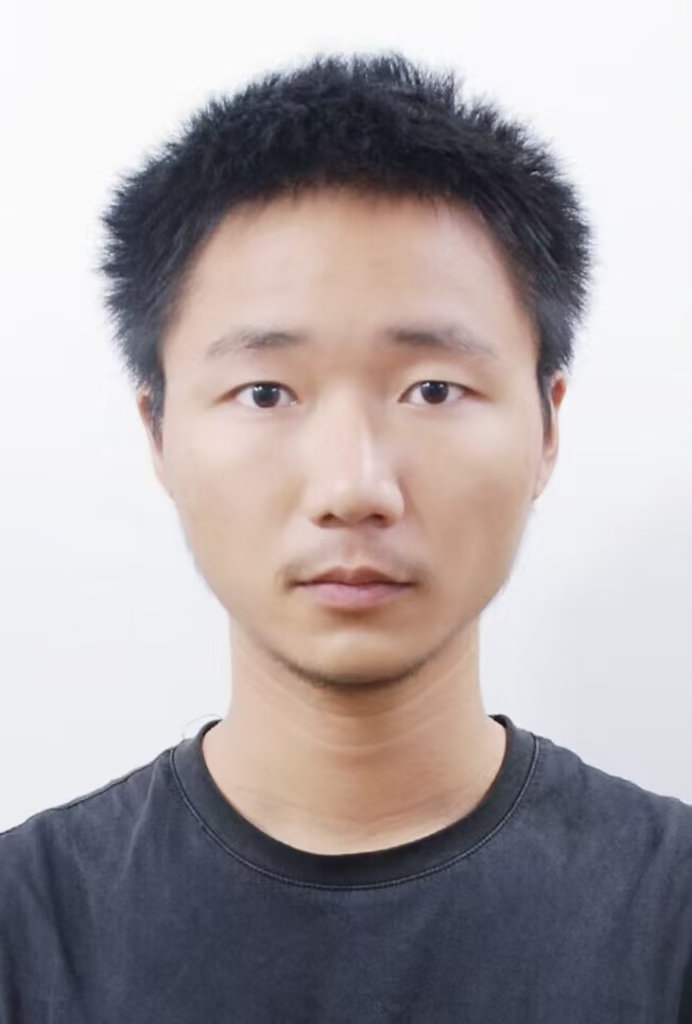}}]
	{Kai Li}
	received the B.S. degree in integrated circuit design and integrated systems from Huazhong University of Science and Technology, Wuhan, China, in 2023, where he is currently pursuing the M.E. degree in integrated circuit science and engineering with the School of Integrated Circuits. His current research interests include high-speed implementation of post-quantum crypto-processor and RISC-V domain-specific architecture.
\end{IEEEbiography}

\vfill

\end{CJK} 
\end{document}